 \documentclass[lettersize,journal]{IEEEtran}
\usepackage{graphicx}
\usepackage{epstopdf}
\usepackage{float}
\usepackage{array}
\usepackage{amsmath}
\usepackage{amssymb}
\usepackage{amsthm}
\usepackage{eqparbox}
\usepackage{fixltx2e}
\usepackage{makecell}
\usepackage{mathrsfs}
\usepackage{amsmath}
\usepackage{multicol}
\usepackage{flushend}

\usepackage{url}
\usepackage{cite}
\usepackage[justification=centering]{caption}
\usepackage{bm}
\usepackage{threeparttable}
\usepackage{subeqnarray}
\usepackage{cases}
\usepackage{color}
\usepackage[lined,boxed,commentsnumbered,ruled]{algorithm2e}
\usepackage{stfloats}
\usepackage[section]{placeins}
\usepackage{setspace}
\ifCLASSOPTIONcompsoc
\usepackage[caption=false,font=normalsize,labelfont=sf,textfont=sf]{subfig}
\else
\usepackage[caption=false,font=footnotesize]{subfig}
\fi

\allowdisplaybreaks[4]

\newcommand{\diag}{\mathop{\mathrm{diag}}}


\begin{document}
	\captionsetup{font={small}}
	%\captionsetup{font={large}}		
	\title{Multi-hop Multi-RIS Wireless Communication Systems: Multi-reflection Path Scheduling \\and Beamforming }
	\author{Xiaoyan Ma,~\IEEEmembership{Member, IEEE,}
		Haixia~Zhang,~\IEEEmembership{Senior Member, IEEE,} 
		Xianhao Chen,~\IEEEmembership{{Member}, IEEE,}\\
		Yuguang Fang,~\IEEEmembership{Fellow, IEEE,}
		and  Dongfeng~Yuan,~\IEEEmembership{Senior Member, IEEE}		

  	\thanks{This work was supported in part by  the Joint Funds of NSFC under Grant No. U22A2003, in part by the Basic and Applied Basic Research Foundation of Guangdong Province under Grant No. 2021B1515120066 and in part by the 
		the National Natural Science Foundation of China under Grant No. 62271288. (\emph{Corresponding author: Haixia Zhang}).
	}
	
	\thanks{X. Ma is with Shandong Key Laboratory of Wireless Communication Technologies, Jinan 250061, China, and School of Information Science and Engineering, Shandong University, Qingdao 266237, China (email: maxiaoyan06@mail.sdu.edu.cn).}
	\thanks{H. Zhang is with Shandong Key Laboratory of Wireless Communication Technologies, Jinan 250061, China, and the School of Control Science and Engineering, Shandong University, Jinan 250061, China (email:haixia.zhang@sdu.edu.cn).}
       \thanks{X. Chen is with the Department of Electrical and Electronic Engineering, the University of Hong Kong (email:xchen@eee.hku.hk).}
	\thanks{Y. Fang is  with the Department of Computer Science, City University of Hong Kong (email: my.Fang@cityu.edu.hk).}	
   \thanks{D. Yuan is with Shandong Key Laboratory of Wireless Communication Technologies, Jinan 250061, China (email: dfyuan@sdu.edu.cn).}
	}	
	\maketitle
	\IEEEpeerreviewmaketitle
	\vspace{-1cm}
	\begin{abstract}
		Reconfigurable intelligent surface (RIS) provides a promising way to proactively augment propagation environments for better transmission performance in wireless communications. Existing multi-RIS works mainly focus on link-level optimization with predetermined transmission paths, which cannot be directly extended to system-level management, since they neither consider the interference caused by undesired scattering of RISs, nor the performance balancing between different transmission paths.
		To address this, we study an innovative multi-hop multi-RIS communication system, where a base station (BS) transmits information to a set of distributed users over multi-RIS configuration space in a multi-hop manner. The signals for each user are subsequently reflected by the selected RISs via multi-reflection line-of-sight (LoS) links. 
		To ensure that all users have fair access to the
		system to avoid excessive number of RISs serving one user, we aim to find the optimal beam reflecting path for each user, while judiciously determining the path scheduling strategies with the corresponding beamforming design to ensure the fairness.
		Due to the presence of interference caused by undesired scattering of RISs, it is highly challenging to solve the formulated multi-RIS multi-path beamforming optimization problem.
		To solve it, we first derive the optimal RISs' phase shifts and the corresponding reflecting path selection for each user based on its practical deployment location. With the optimized multi-reflection paths, we obtain a feasible user grouping pattern for effective interference mitigation by constructing the maximum independent sets (MISs).
	    Finally, we propose a joint heuristic algorithm to iteratively update the beamforming vectors and the group scheduling policies to maximize the minimum equivalent data rate of all users. Numerical results demonstrate that the proposed transmission framework achieves superior throughput performance than benchmark schemes. Useful insights on how to leverage multi-reflection paths over RISs to boost the throughput performance are also drawn under different settings for the multi-hop multi-RIS communication systems.
	\end{abstract}	
	
	
\vspace{-0.8cm}
	\section{Introduction}
	\IEEEPARstart{T}{raditional} communication systems mainly rely on transmission technologies at transmitting and/or receiving ends, like channel coding, modulation and power control, to adapt to the propagation environments for better performance \cite{Channel_coding,modulation1,modulation}.
	All these technologies should be carefully designed to eliminate the distortion brought by the transmission environments passively \cite{LAN}. 
	Recently, reconfigurable intelligent surface (RIS) has been proposed as a promising technology to proactively manipulate and augment transmission environments to enhance the performance of communication systems \cite{LAN11,Di1122}.
	RIS can be viewed as a planar surface, which comprises a large number of low-cost passive reflecting elements. Each element can be digitally controlled to induce an amplitude change and/or phase shift to the incident signals, thereby collaboratively altering physical transmission environments. Besides, from the implementation perspective, RISs possess appealing features such as low cost and lightweight, thus can be easily and conveniently installed in the environment, for example, mounted on walls or ceilings \cite{MA2}. Besides, since RISs are complementary devices, deploying them in the existing wireless systems only needs minor modifications of the communication protocols, and no other significant changes on standards and hardware are required. Furthermore, RISs are usually
	much cheaper than active small base stations (SBSs)/relays, therefore can be deployed easily and rapidly \cite{Huang1}. Finally, due to its flexible form factor, a surface can be made and deployed without damaging the environmental aesthetic beauty, rather improving living environment if being carefully designed \cite{LAN,LAN11}.
	
	Due to these attractive features of RIS, it has been  intensively studied in combination with various key communication technologies, such as orthogonal frequency division multiplexing (OFDM) \cite{OFDM1,OFDM2,OFDM3}, non-orthogonal multiple access (NOMA) \cite{NOMA1,NOMA2,NOMA3}, simultaneous wireless information and power transfer (SWIPT) \cite{SWIPT1,SWIPT2}, mobile edge computing (MEC) \cite{MEC1,MEC2,MEC3} and physical layer security \cite{Li1,zhang1,Sun1}. Existing works mainly consider one or multiple distributed RISs,
	where each RIS individually reflects the signals to users, and there is no cooperation among multiple RISs in most of the existing works. This simplified approach generally leads to sub-optimal performance, since those multi-reflection paths that may offer high cooperation beamforming gains are not well utilized. In addition, multi-hop reflections established through inter-RIS links can provide  additional channel diversity, bypassing potential obstacles that cannot be circumvented by single RIS, thus forming blockage-free paths even under harsh radio environments \cite{Zheng1}.
	
	Inspired by the advantages of cooperative transmission in multi-hop multi-RIS systems, Han {\sl et al.} \cite{double} investigated double-RIS-assisted single-user wireless communication systems. As an early work, it is shown in \cite{double} that by cooperatively designing the phase shifts, a double-reflection path can provide passive beamforming gain that increases quartically with the total number of RISs' reflecting elements, which significantly outperforms the quadratic growth of the single-RIS case with the same number of reflecting elements. This work has been extended to multi-user systems \cite{R1,R6,MA1}, where the superiority of utilizing multi-RIS multi-hop transmissions are further confirmed.
	
	The above works on multi-RIS transmission mainly focus on link-level optimization, i.e., the transmission path for each user is predetermined. For a generalized communication system with more available RISs and randomly distributed users, different line-of-sight (LoS) links can be created for each user via multi-hop reflections. Mei and Zhang \cite{Mei1} have made an initial attempt to address this issue. They consider downlink multi-hop scenarios aided by multiple RISs, where the ideal interference-free transmission environments are considered. With this interference-free assumption, \cite{Mei1} directly simplifies the maximize minimum (max-min) data rate problem to the max-min channel gain problem. Such actions severely limit the overall resource utilization. Besides, as shown in \cite{interference1} and \cite{interference2}, the interference is an increasingly challenging issue in RIS-assisted wireless communications. Since passive RISs do not have signal processing abilities and can only reflect the incident signals, if the interference coordination schemes are not carefully designed, the cumulated interference caused by the undifferentiated reflections of RISs will mess up the radio transmission environments, and finally degrade the system performance.

	\begin{figure}[t]
		\vspace{-0.5cm}
		\setlength{\belowcaptionskip}{-0.6cm}
		\centering
		\includegraphics[width=0.45\textwidth]{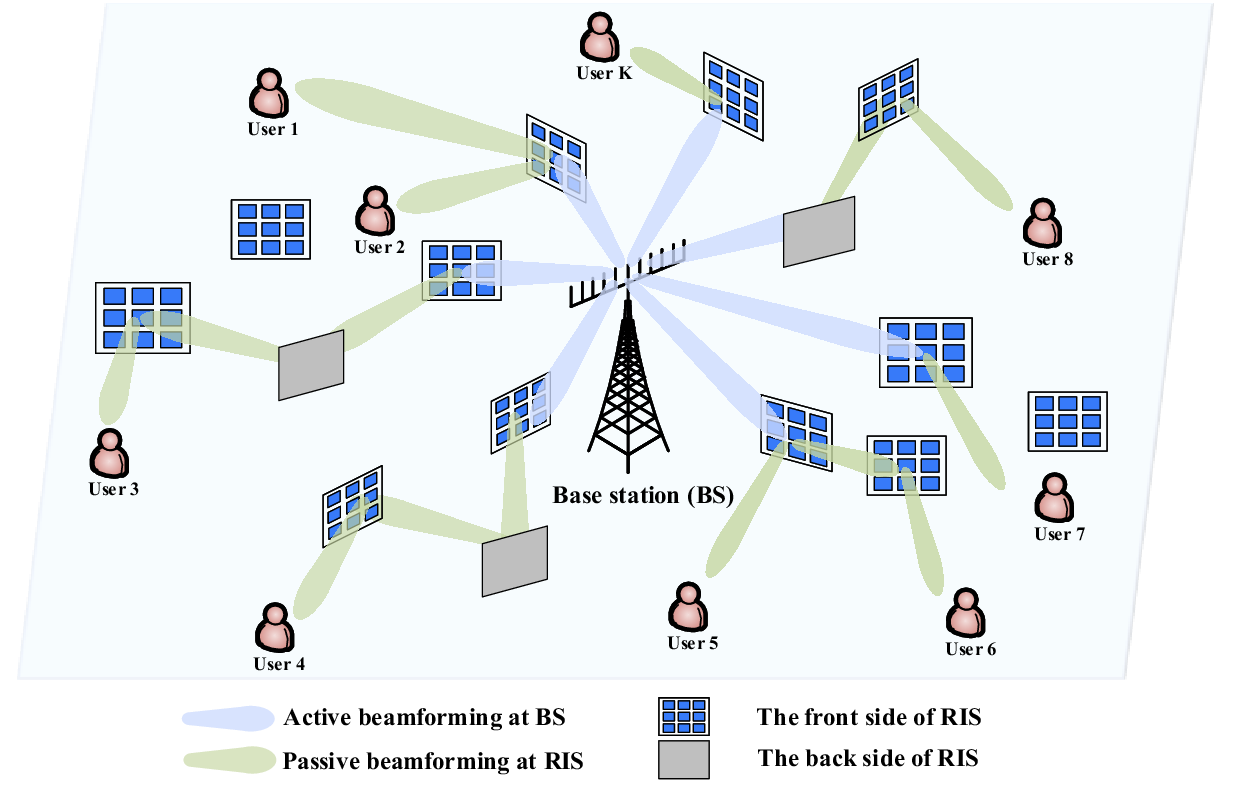}\\
		\caption{The considered multi-hop multi-RIS communication system.}
		\label{fig:channel_model}
	\end{figure}

	To address the interference management in multi-RIS assisted communication systems, in this paper, we study a multi-hop multi-RIS  cooperative  communication scenario, where a BS equipped with multiple antennas transmits signals to a set of randomly distributed users, as shown in Fig. \ref{fig:channel_model}. Here, multiple RISs are carefully pre-placed in  proper locations to create LoS transmission opportunities between BS and all the users. 
	To ensure that all users have fair access to the system  to avoid excessive number of RISs serving one user, we first find the optimal beam reflective path for each user, along with the optimal phase shifts of the selected RISs. Afterwards, to mitigate inter-user interference caused by undesired scatterings of RISs, we divide users into different activation groups by constructing the maximum independent sets (MISs). Finally, we design a group scheduling strategy and the corresponding beamforming optimization policy at the BS to maximize the minimum equivalent data rate of all users to ensure the fairness. The main contributions are summarized as follows.
	
	\begin{itemize}
		\item We design a novel framework for multi-hop multi-RIS communication systems, where a BS transmits signals to a set of distributed users through multi-hop reflections. Different from existing works with predetermined transmission paths, we explore the characteristics of RIS-assisted multi-reflection channels, select the multi-hop transmission path with the maximum equivalent channel gain for each user. Further, to avoid the inter-path interference caused by the undifferentiated reflections of RISs, we propose a new path-separation constraint, with which only the paths that do not interfere with each other can be activated at the same time, thereby guarantee the controllability of radio transmission environments.
	
	    \item To find the optimal multi-hop reflection path for each user, we analyzing the characteristics of the RIS-assisted multi-reflection channels and reformulating the channels based on the composition features. By doing so, we derive the optimal RISs' phase shift design and assign corresponding weight to each link (i.e., BS-RIS links, inter-RIS links and RIS-user links). With defined link weight,  we recast the multi-hop transmission design as a shortest-path finding problem in graph theory, and select the optimal transmission path with the maximum equivalent channel gain for each user.
	    
	    \item We adopt conflict graph to model the inter-path interference, and divide users into different activation groups to satisfy the path-separation constraint. Specifically, in the modeled conflict graph, each vertex represents one multi-hop path. For any two vertices, if they cannot satisfy the path-separation constraint, an edge is built between them. Based on the conflict graph, we introduce the concept of maximal independent sets (MISs), which equivalently represents the maximum number of paths that can transmit simultaneously without inter-path interference. And we divide users into different activation groups to mitigate inter-path interference with the help of MISs. During the scheduling period, all the activation groups take turns to access the base station (BS) for information transmission, and we propose efficient scheduling policy to guarantee the fairness among different activation groups.	
	
		\item
			To ensure that each user has fair access to the network, the beamforming pattern at the BS is optimized for each activation group such that the minimum equivalent data rate of all users is maximized. In particular, we introduce a series of auxiliary variables to transform the max-min equivalent data rate problem into a feasibility check problem, which can be solved by classical convex optimization approaches. Numerical results demonstrate that the proposed transmission framework achieves higher throughput than the benchmark schemes. Useful insights on how to obtain the optimal multi-reflection cooperative transmission designs are also drawn under different settings of the multi-hop multi-RIS systems.
	\end{itemize}
	
	The remainder of this paper is organized as follows. We first present the system model and transmission scheme in Section II. In Section III, we formulate the optimization problem to maximize the minimum data rate of all users. In Section IV, we describe the proposed three-step framework for multi-hop multi-RIS transmission systems. In Section V, we provide simulation results to demonstrate the superiority of the proposed cooperative multi-hop multi-RIS transmission framework. Finally, we conclude this paper in Section VI.
	
	The notations are listed as follows.
	Bold symbols in capital letter and small letter denote matrices and vectors, respectively.
	$\mathcal{CN}(\mu,\sigma^{2})$ denotes the circularly symmetric complex Gaussian (CSCG) distribution with mean $\mu$ and variance $\sigma^{2}$.
	$\text{Tr}(\mathbf{A})$  represents the trace of matrix $\mathbf{A}$.
	$\mathbf{A}=\diag(\bm{a})$ means that $\mathbf{A}$ is the diagonal matrix of the vector $\bm{a}$. $A_{i,j}$ represents the element at row $i$, column $j$ of matrix $\mathbf{A} $ and $a_{n}$ represents the $n$-th elements of vector $\bm{a}$. 
	$\otimes$ is the Kronecker product.
	$||\mathbf{w}||$ denotes the Euclidean norm.
	$\mathbf{G}^{T}$ and $\mathbf{G}^{H}$ denote the transpose and conjugate transpose of matrix $\mathbf{G}$, respectively. 
	For complex variable $x$, $\angle x$ represents its phase angle.
	$\mathcal{H}_{M}^{+}$ represents the set for Hermitian positive semi-definite matrices of dimension $M \times M$.

	\section{System Model and Transmission Schematic Design}
	In this section, we first introduce the basic model of the multi-hop multi-RIS transmission system. Then, we provide the channel models between BS-RIS, inter-RIS and RIS-user. Finally, we present the transmission schemtic design of the considered multi-hop multi-RIS cooperation transmission system. 
	\vspace{-0.5cm}
	\subsection{System Model}
	As shown in Fig. \ref{fig:channel_model}, we consider a downlink multi-user communication system, where $J$ distributed RISs are carefully pre-deployed to assist signal transmissions between the BS and $K$ single-antenna users. The users are randomly distributed within the area of interest. Assume that the BS is equipped with $M_0$ antennas, while each RIS has $M_j$ reflecting elements, $j=1,2,...,J$. Moreover, each RIS can only achieve half-space reflection, i.e., only the signal incident on its reflection side can be reflected. Thus, for any two nodes $i$ and $j$, if they are both RISs, each of them needs to be located in the reflection half-space to achieve effective signal reflection.
	We assume that the passive reflecting elements of each RIS are arranged in a uniform rectangular array (URA) perpendicular to the ground and facing a fixed direction, while the BS employs the uniform linear array (ULA) antenna configuration. The antenna and element spacings at the BS and each RIS are assumed to be $d_B$ and $d_R$, respectively. 
	To enhance the received signal strength at users, we proactively design the system to transmit messages only through the virtual LoS paths established via multi-reflection. Furthermore, to mitigate the severe inter-path interference caused by the undesired scattering of RISs, the signal paths for all $K$ users should be sufficiently separated and thus each RIS is associated with at most one user at a time. However, each RIS can serve multiple users over different time via proper scheduling policy. 
	
	We denote the sets of RISs and users as $\mathcal{J}=\lbrace 1,2,...,J \rbrace$ and  $\mathcal{K}=\lbrace 1,2,...,K \rbrace$, respectively. To clearly represent the nodes, i.e., BS, RISs and users, in the system, we refer to BS as node $0$, RIS $1$ to RIS $J$ as node $1$ to node $J$, user $1$ to user $K$ as node $J+1$ to node $J+K$, thereby there are a total of $1+J+K$ nodes in the system, and we use  $d_{i,j}$, $i \neq j$, $i \in \lbrace 0,1,2,...,J \rbrace$, $j \in \lbrace 1,2,...,J, J+1,...,J+K \rbrace$ to denote the distance between node $i$ and $j$. 
	%To ensure the far-field propagation between any two nodes, we assume that $d_{i,j} \geq d_0$, where $d_0$
	%denotes the minimum distance to guarantee this condition. According to \cite{Mei1}, $d_0$ should satisfy $d_0 \gg \frac{\sqrt{M_i}d^2}{\lambda}$,  where $M_i$ with $i=0,1,2,...,J$ represents the number of antennas at the BS or the number of reflecting elements at RIS $j$, $d\in \lbrace d_B, d_R \rbrace$ is the antenna or element spacing at the BS or RIS, and $\lambda$ denotes the wavelength. 
	By carefully selecting RISs, LoS dominant transmission can be achieved between the BS and each user, and to enhance the signal strength of the multi-hop transmissions, we only exploit LoS transmissions in the system. Moreover, to describe the communication condition between any two nodes $i$ (BS/RIS) and $j$ (RIS/user) in the considered system, we define
	a binary indicator $l(i,j) \in \lbrace 0,1 \rbrace$. In particular,  $l(i,j)=1$ indicates that nodes $i$ and $j$ can communicate with each other, and $l(i,j)=0$ otherwise. In addition, we set $l(i,i) = 0, \forall \, i$ and have $l(i,j)=l(j,i), \forall \, i,j$.
	These binary indicators $l(i,j)$  can be viewed as prior information once RISs are carefully deployed. Based on these known communication conditions, specific multi-reflection path can be established for each user by properly selecting a subset of RISs.
	For example, if $l(0,i)=l(i,j)=l(j,J+k)=1$, $i,j \in \mathcal{J}$, $k \in \mathcal{K}$, this indicates that the signals transmitted from BS can be successively reflected by RISs $i$ and $j$ to reach user $k$. Thus, we can select RISs $i$ and $j$ as the associated RISs for user $k$. For RISs that are not associated with any user, the BS can turn them off via control links so as to reduce the scattered interference in the system \cite{Mei1,R2}. 
	
\subsection{Channel Model and Transmission Scheme}
	We define $\mathbf{H}_{0,j} \in \mathbb{C}^{M_j \times M_0}$ as the channel from BS to RIS $j$, $\bm{g}_{j,J+k}^H \in \mathbb{C}^{1 \times M_j}$ as the channel from RIS $j$ to user $k$, and $\mathbf{S}_{i,j} \in \mathbb{C}^{M_j \times M_i}$, $i \neq j, \, i,j \in \mathcal{J}$ as the channel from RIS $i$ to RIS $j$.
	The numbers of reflecting elements in each RIS's horizontal and vertical directions are assumed to be $M_{jx}$ and $M_{jy}$ with $M_{jx}M_{jy}=M_j$. 
	The reflection coefficient matrix of each RIS $j, \, j \in \mathcal{J}$ is given by $\mathbf{\Phi}_{j} \in \mathbb{C}^{M_j \times M_j} $, which can be further expressed as $\mathbf{\Phi}_{j}= \diag (\pmb{\phi}_{j})$. $\pmb{\phi}_j=[\phi_{j,1},\phi_{j,2},...,\phi_{j,M_{j}}]^T$ is the corresponding reflection vector of the $j$-th RIS, where
	$\phi_{j,m}=\beta_{j,m}e^{j\theta_{j,m}}$, $m = 1,2,...,M_{j}$. Moreover, $\theta_{j,m} \in [0,2\pi)$ and $\beta_{j,m} \in [0,1]$ respectively represent the phase and amplitude change brought by the $m$-th reflecting element of RIS $j$ to the incident signals.
	Without loss of generality, we set $\beta_{j,m} =1$. If RIS $j$ is selected to assist the signal transmission, all the elements of the RIS $j$ are switched on to fully reflect the incident signals. 
	
	Next, we characterize the LoS channel between two nodes in the system, which is modeled as the product of array responses of the transmitter and receiver \cite{R3,R4,R5,R7,ML1}. For convenience, we define the following steering vector
	\begin{equation}
	\setlength{\abovedisplayskip}{2pt}
	\setlength{\belowdisplayskip}{2pt}
		\begin{aligned} 
			\bm{e}(\delta,N)=[1,e^{-j\pi\delta},e^{-j2\pi\delta},...,e^{-j(N-1)\pi\delta}] \in \mathbb{C}^{N \times 1},
		\end{aligned}
	\end{equation}
	where $N$ denotes the number of elements in a ULA and $\delta$ denotes the phase difference between the observations at two adjacent elements. Then, the array response at the BS is expressed as 
	\begin{equation}
	\setlength{\abovedisplayskip}{2pt}
	\setlength{\belowdisplayskip}{2pt}
		\begin{aligned} 
			\bm{\alpha}_B(\vartheta)=\bm{e}\left(\frac{d_B}{\lambda}\sin\vartheta, M_0\right),
		\end{aligned}
	\end{equation}	
	where $\vartheta$ denotes the angle-of-departure (AoD) related to the BS antenna boresight. For the URA at each RIS, its array response is expressed as the Kronecker product of two steering vectors in the horizontal and vertical directions, respectively, i.e.,
	\begin{equation}
		\begin{aligned} \label{arrayR1}
			\bm{\alpha}_{R,j}(\vartheta^a,\vartheta^e)= &\, \bm{e}\left(\frac{d_R}{\lambda}\sin\vartheta^a \cos \vartheta^e,M_{j,x}\right)\\
			 &\otimes 
			\bm{e}\left(\frac{d_R}{\lambda}\sin\vartheta^a \sin \vartheta^e,M_{j,y}\right).
		\end{aligned}
	\end{equation}

	In \eqref{arrayR1}, $\vartheta^a$ and $\vartheta^e$ denote the elevation and azimuth angle parameters, respectively. With these expressions for array responses, we define $\vartheta_{0,j}$ as the AoD from the BS to RIS $j$,  $\vartheta^a$ and $\vartheta^e$ as the azimuth and elevation AoD from RIS $i$
	to node $j$ (RIS or user), respectively, $\varphi^a$ and $\varphi^e$ as the azimuth and elevation angle-of-arrival (AoA) at RIS $j$ from node $i$ (BS or RIS), respectively. 
	Based on the above definitions, $l(0,j)=1$ means that BS can communicate with RIS $j$ through LoS channel. The channel from BS to RIS $j$ can be expressed as
	\begin{equation}
	\setlength{\abovedisplayskip}{2pt}
	\setlength{\belowdisplayskip}{2pt}
		\begin{aligned} \label{H}
			\mathbf{H}_{0,j}=\sqrt{\beta_H} \tilde{\bm{h}}_{0,j,2} \tilde{\bm{h}}^H_{0,j,1}, \,\,\,\, j \in \mathcal{J},
		\end{aligned}
	\end{equation}
	where $\beta_H=\beta_0/d^2_{0,j}$ denotes the free-space path-loss with $\beta_0$ represents the channel gain at the reference
	distance of $1$ meter, 
	$\tilde{\bm{h}}_{0,j,2}=\bm{\alpha}_{R,j}(\varphi^a_{0,j},\varphi^e_{0,j})$ represents the receiving array response of RIS $j$ and $\tilde{\bm{h}}_{0,j,1}=\bm{\alpha}_B(\vartheta_{0,j})$ is the transmitting array response of the BS. If $l(i,j)=1, \, i,j \in \mathcal{J}$, RIS $i$ can communicate with RIS $j$ through LoS channel, and the channel from RIS $i$ to RIS $j$ can be expressed as
	\begin{equation}
	\setlength{\abovedisplayskip}{2pt}
	\setlength{\belowdisplayskip}{2pt}
		\begin{aligned} \label{S}
			\mathbf{S}_{i,j}=\sqrt{\beta_{S_{i,j}}} \, \tilde{\bm{s}}_{i,j,2} \tilde{\bm{s}}^H_{i,j,1}, \,\, i,j \in \mathcal{J},\, i\neq j.
		\end{aligned}
	\end{equation}
	In \eqref{S}, $\beta_S=\beta_0/d^2_{i,j}$ is the corresponding free-space path-loss, $\tilde{\bm{s}}_{i,j,2}= \bm{\alpha}_{R,j}(\varphi^a_{i,j},\varphi^e_{i,j})$ is the receiving array response at RIS $j$ and $\tilde{\bm{s}}_{i,j,1}=\bm{\alpha}_{R,i}(\vartheta^a_{i,j},\vartheta^e_{i,j})$ is the transmitting array response at RIS $i$. Similarly, the channel from RIS $j$ to user $k$ when $l(j,J+k)=1$ can be expressed as
	\begin{equation}
	\setlength{\abovedisplayskip}{2pt}
	\setlength{\belowdisplayskip}{2pt}
		\begin{aligned} \label{g}
			\bm{g}_{j,J+k}^H=\sqrt{\beta_{g_{j,J+k}}} \, \tilde{\bm{g}}^H_{j,J+k}, \, \, \, j\in \mathcal{J}, k\in\mathcal{K},
		\end{aligned}
	\end{equation}
	where $\beta_{g_{j,J+k}}=\beta_0/d^2_{j,J+k}$ represents the corresponding free-space path-loss, and $\tilde{\bm{g}}_{j,J+k}=\bm{\alpha}_{R,j}(\vartheta^a_{j,J+k},\vartheta^e_{j,J+k})$ denotes the transmit array response at RIS $j$. The above AoAs and AoDs of the BS-RIS and inter-RIS channels are assumed to be known, since once the BS and RISs are deployed, these angle information can be directly obtained from their geographical locations \cite{Di1122}. Therefore, for the system design, we only need to obtain the CSI between the last RIS in the multi-reflection path and its target user. 
	Considering that RISs are usually deployed at the places that can establish strong LoS links to users, the location based channel information can be directly used to assist the system design \cite{location1,location}, and the location information of users can be obtained through the global positioning system (GPS). Besides, some advanced channel estimation methods, i.e., anchor-assisted method, have been proposed to perform efficient channel estimation for RIS-assisted systems \cite{CEEE1,CEEE2,Anchor1,Anchor2}. For indoor environment, we can leverage existing wireless local area network (WLAN) infrastructures for indoor positioning,  or use Bluetooth to track users. Besides, many indoor environments have security cameras, we can also use image based vision positioning techniques \cite{Indoor1, Indoor2}. Therefore, in this paper, we assume that all the CSIs needed are perfectly known, and we focus on the framework design for the mult-hop multi-RIS communication systems. 
	
	Based on \eqref{H} - \eqref{g}, we can characterize the multi-reflection LoS channel between the BS and user $k$ with given multi-reflection path. Specifically, we use 
	$\Omega(k)=\lbrace  a_1^{(k)},a_2^{(k)},...,a_{N_k}^{(k)} \rbrace$ to
	denote the multi-reflection path from the BS to user $k$, where $N_k \geq 1$ denotes the number of associated RISs for user $k$, $a_n^{(k)}, n \in \mathcal{N}_k \overset{\vartriangle}{=} \lbrace 1,2,...,N_k \rbrace$ represents the index of the $n$-th associated RIS. For convenience, we define $a_0^{(k)}=0$ and $a_{N_k+1}^{(k)}=J+k$ to denote the BS and the target user $k$, respectively. Then, to ensure that each RIS in $\mathcal{N}_k$ only reflects user $k$'s signal at most
	once, the following constraints should be met
	\begin{equation}
	\setlength{\abovedisplayskip}{2pt}
	\setlength{\belowdisplayskip}{2pt}
		\begin{aligned} \label{C1}
			a_n^{(k)} \in \mathcal{J}, \,\,\, a_n^{(k)} \neq a_{n^{'}}^{(k)}, \,\,\,
			\forall n,n^{'} \in \mathcal{N}_k, \,\, n\neq n^{'}, \,\,k\in \mathcal{K}.
		\end{aligned}
	\end{equation}
Moreover, the consecutive nodes in $\Omega(k)$ should be able to communicate through an LoS channel, i.e., 
	\begin{equation}
	\setlength{\abovedisplayskip}{2pt}
	\setlength{\belowdisplayskip}{2pt}
		\begin{aligned} \label{C2}
			l\left(a_n^{(k)},a_{n+1}^{(k)}\right)=1, \,\,\, \forall n \in \mathcal{N}_k \,	\cup \lbrace 0 \rbrace, \,\, k \in \mathcal{K}.	
		\end{aligned}
	\end{equation}
Furthermore, to avoid the scattered inter-user interference, we need to ensure that there is no direct communication link (i.e., the transmission link is blocked by obstacles or exceeds the communication distance) between any two nodes belonging to different selected paths except for the common starting node $0$ (BS). Thus, we introduce the new path separation constraint, i.e.,
	\begin{equation}
	\setlength{\abovedisplayskip}{2pt}
	\setlength{\belowdisplayskip}{2pt}
		\begin{aligned} \label{C3}
			l\left(a_n^{(k)},a_{n^{'}}^{(k^{'})}\right)=0, \,\, a_n^{(k)}\neq &a_{n^{'}}^{(k^{'})}, \\
			&\forall \,n,n^{'} \neq 0,\,\, k,k^{'} \in \mathcal{K},\,\, k\neq k^{'}.
		\end{aligned}
	\end{equation}
	Note that the condition $a_n^{(k)}\neq a_{n^{'}}^{(k^{'})}$ ensures that there is
	no common node (except for the BS) between any two reflection paths $\Omega(k)$ and $\Omega(k^{'})$ for user $k$ and $k^{'}$, respectively. With the above definitions, a reflection path $\Omega(k)$ is feasible if and only if the constraints in \eqref{C1} - \eqref{C3} are satisfied. It is worth noting that if the number of users is small and their distribution is sparse, constraint \eqref{C3} can be easily satisfied as observed in \cite{Mei1}. However, in practical scenarios, the number of served users can be large and their locations can be very close to each other. In order to meet the path separation constraint, we need to divide users into different activation groups, only a subset of users that satisfy \eqref{C1} - \eqref{C3} can be activated at the same time. Thereby, constraint \eqref{C3} is further modified into
	\begin{equation}
	\setlength{\abovedisplayskip}{2pt}
	\setlength{\belowdisplayskip}{2pt}
		\begin{aligned} \label{C4}
			l^{\,q}\left(a_n^{(k)},a_{n^{'}}^{(k^{'})}\right)=0, \,\, &a_n^{(k)}\neq a_{n^{'}}^{(k^{'})}, \\ &\forall \,n,n^{'} \neq 0,\,\, k,k^{'} \in \mathcal{K},\,\, k\neq k^{'},
		\end{aligned}
	\end{equation}
where $q=1,2,...,Q$ is the index of the time slot with $Q$ being the total number of time slots that the whole scheduling time is divided. We use $t_q$ to represent the proportion of time allocated to time slot $q$, we have $\sum_{q=1,2,...,Q} t_q=1 $. Constraint \eqref{C4} means that paths that do not interfere with each other can be active at the same time. How to construct the user activation groups and determine the time allocation parameter $t_q$, $q=1,2,...,Q$ is further introduced in Section IV.

There are two main advantages of the considered transmission scheme. First, one RIS can only serve one user at a time, which means that in any transmission path, there is only one target receiver for each transmitter except for the BS. {For example, for user $1$, there is a multi-reflection path $\Omega(1)=\lbrace 1,5,10 \rbrace$. This means that the BS can transmit messages to user $1$ through the multi-hop LoS path established by RISs $1$, $5$ and $10$.} During the inter-RIS transmissions,  the signal transmission pattern is one-to-one, i.e., RIS $1$ only needs to target RIS $5$ for signal delivery. In this case, the phase shift optimizations of RISs are very simple, it only needs to achieve the phase alignment with the channel between RIS $1$ and RIS $5$ \cite{double}. Since all the RISs are carefully pre-placed, the channel state information of all the inter-RIS links can be easily obtained. Thus we can design the optimal phase shift pattern between any two RISs in advance, which can greatly reduce the design complexity caused by the real-time phase shift design. The second advantage comes from the path separation constraints shown in \eqref{C4}. If the BS transmits signals to all users at the same time, in addition to the designed multi-reflection LoS transmission paths, there also exists undesired scattering of RISs. This kind of scattered interference will mess up the transmission environment, cause serious inter-path interference, and finally decrease the system performance. To avoid this, we introduce the path-separation constraint \eqref{C4} and divide users into different activation groups to eliminate the inter-path inference and guarantee the controllability of the radio environment.

\section{Problem Formulation for the Multi-hop Multi-RIS Cooperative Transmission System}

%In this section, the optimal configurations of RISs and the corresponding link weights are first defined in order to select the proper multi-reflection paths for each user. Then, based on the designed paths, we are able to formulate the joint beamforming and user scheduling problem to  maximize the minimum equivalent data rate of all users.
In this section, we formulate an optimization problem maximizing the minimum equivalent data rate of all users to guarantee the fairness.
We first model the equivalent channel from the BS to user $k$ with given multi-reflection path $\Omega(k)$, which can be expressed as 
	\begin{equation}
		\begin{aligned} \label{Equ_Channel}
			&\bm{h}_{0,J+k}\left(\Omega(k)\right)= \\&\bm{g}_{a_{N_k}^{(k)},J+k}^H \mathbf{\Phi}_{a_{N_k}^{(k)}}
			\left( \prod_{n \in \mathcal{N}_k,n\neq N_k} \!\!\!\!\!\!\!\! \mathbf{S}_{a_n^{(k)},a_{n+1}^{k}} \mathbf{\Phi}_{a_n^{(k)}}
			\right) \mathbf{H}_{0,a_1^{(k)}}.
		\end{aligned}
	\end{equation}
Obviously, it depends on the $N_k$ selected RISs and the corresponding phase shift design.
	As discussed before, to avoid the serious inter-path interference caused by the simultaneous transmissions of all users, we divide all the users into different activation groups, only a subset of users that satisfy \eqref{C1} - \eqref{C3} can be activated at any given time. For one specific time slot $q$, $t_q$ denotes the fraction of time allocated to it, and $x_k^{q} $ denotes the transmitted signal from the BS to user $k$, with $\mathbb{E} \lbrace |x_k^{q}|^2 \rbrace =1 $. The transmitted signals for all users at time slot $q$ from the BS is expressed as $\bm{x}^{q}=\sum \bm{w}_k^{q} x_k^{q} $, where $\bm{w}_k^{q} \in \mathbb{C}^{M_0 \times 1}$ is the corresponding transmit beamforming vector for user $k$ at time slot $q$, $q=1,2,...,Q$. With a given reflection path $\Omega(k)=\left\{a_{1}^{(k)}, a_{2}^{(k)}, \cdots, a_{N_{k}}^{(k)}\right\}, k \in \mathcal{K} $, the received signal at user $k$  \cite{Zhang_HX}is expressed as 
	\begin{equation}
	\setlength{\abovedisplayskip}{2pt}
	\setlength{\belowdisplayskip}{2pt}
		\begin{aligned}
			y_k^{q}=\bm{h}_{0,J+k}\left(\Omega(k)\right)\bm{x}^{q}+n_k,
		\end{aligned}
	\end{equation}  
where $n_{k} \sim \mathcal{CN}(0,\sigma_{0}^{2})$ denotes the complex additive white Gaussian noise (AWGN) at the $k$-th user. 
The corresponding signal to interference plus noise ratio (SINR) for user $k$ at time slot $q$ can be expressed as
	\begin{equation} \label{SINR_K_R}
	\setlength{\abovedisplayskip}{2pt}
	\setlength{\belowdisplayskip}{2pt}
		\begin{aligned}
			\gamma_k^{q}=\frac{|\bm{h}_{0,J+k}\left(\Omega(k)\right)\bm{w}_k^{q}|^2}
			{
				\sum_{u\in \mathcal{I}_{q},u \neq k}|\bm{h}_{0,J+k}\left(\Omega(k)\right)\bm{w}_u^{q}|^2+\sigma_0^2
			},
		\end{aligned}
	\end{equation}
	where $\mathcal{I}_{q} $ denotes the set of users that served by the BS simultaneously at time slot $q$. As discussed above, to guarantee the controllability of the radio environment, we need to avoid the interference caused by undesired scattering of RISs, thereby we divide users into different activation groups. Since one user may be included in different activation groups, the equivalent data rate of user $k$ during the given scheduling period is the summation of all the time slots that it belongs to, which is expressed as
	\begin{equation}
	\setlength{\abovedisplayskip}{2pt}
	\setlength{\belowdisplayskip}{2pt}
		\begin{aligned}
			C_k=\sum_{k \in \mathcal{I}_{q}} t_q\log(1+\gamma_k^{q}) .
		\end{aligned}
	\end{equation}
	
	To guarantee that all users have access to the network fairly, we aim to maximize the minimum equivalent data rate of all the users. The optimization problem is formulated as
	%\begin{subequations}
		\begin{align}
			\textbf{(P1)}  &\max_{\Omega(k),\mathcal{I}_{q},t_q,\bm{w}_k^{q}} \, \min_{k \in \mathcal{K}} \,\,\, C_k=\sum_{k \in \mathcal{I}_{q}} t_q\log(1+\gamma_k^{q}),	\\ 
			&  \,\,\,\, s.t.   \,\,\,\,
			l\left(a_n^{(k)},a_{n+1}^{(k)}\right)=1, \,\,\, \forall n \in \mathcal{N}_k \,	\cup \lbrace 0 \rbrace, \,\, k \in \mathcal{K}, \label{C11} \\ 
			& \,\,\,\,\,\,\,\,
			a_n^{(k)} \in \mathcal{J}, \,\,\, a_n^{(k)} \neq a_{n^{'}}^{(k)}, \,\,\,
			\forall n,n^{'} \in \mathcal{N}_k, \,\, n\neq n^{'}, \,\,k\in \mathcal{K},\label{C21}\\
			&\,\,\,\,\,\,\,\, 
			 l^{\,q}\left(a_n^{(k)},a_{n^{'}}^{(k^{'})}\right)=0, \,\, a_n^{(k)}\neq a_{n^{'}}^{(k^{'})}, \nonumber \\
			 &\quad\quad\quad\quad\quad\quad\quad
			  \forall \,n,n^{'} \neq 0,\,\, k,k^{'} \in \mathcal{K},\,\, k\neq k^{'},\label{C31} \\
			&  \,\,\,\,\,\,\,\,  0 \leq t_q \leq 1, \sum_{q=1,2,...,Q} t_q=1,\label{C41}\\ 
			& \,\,\,\,\,\,\,\,  \sum_{k \in \mathcal{I}_{q}} ||\bm{w}_k^{q}||^2 \leq P_T\label{C51},
		\end{align}
	%\end{subequations}
	where $P_T$ represents the maximal transmit power at the BS.
	The rest optimization variables in $\textbf{(P1)}$ are listed as follows: 
	\begin{itemize}
	\item
	$\Omega(k)$ represents the designed multi-reflection path for each user $k$, $k\in \mathcal{K}$, which is related to the $N_k$ selected RISs for user $k$ and their corresponding phase shift design.
	
	\item
	$\mathcal{I}_{q}$, $q=1,2,...,Q$ is the designed user activation group. As discussed above,
	to avoid the undesired scattering of RISs, $K$ users are divided into $Q$ activation groups, which is represented as $\mathcal{I}_{1}, \mathcal{I}_{2},...,\mathcal{I}_{Q}$. %The selected multi-reflection paths of all users in the same activation group must satisfy the path separation constraint \eqref{C31}.
	
   \item
    $t_q$, $q=1,2,...,Q$ is the fraction of time allocated to the activation group  $\mathcal{I}_{q}$. During the scheduled period, all the activation groups are scheduled to access the BS, thereby the whole scheduling time is divided into $Q$ time slots and the following constraints must be satisfied $0 \leq t_q \leq 1, \sum_{q=1,2,...,Q} t_q=1$.
    
    \item
    $\bm{w}_k^{q}$ is the the beamforming vector at the BS for user $k$ in time slot $q$.
	\end{itemize}

	For the constraints, \eqref{C11} guarantees the feasibility of the selected links, i.e., the consecutive nodes in $\Omega(k)$  should be able to communicate with LoS channels, \eqref{C21} ensures that each RIS in $\mathcal{N}_k$ can only reflect once, so there is no loop in the multi-reflection path,
	\eqref{C31} avoids the inter-path interference caused by the undesired scattering of RISs, which means that the selected paths for users can only be active at any given time if they do not interfere with each other,
	 \eqref{C41} is the activation group time allocation constraint, 
	  and \eqref{C51} limits the BS' transmit power.

	  The formulated optimization problem $\textbf{(P1)}$ is hard to solve since it includes the multi-reflection path selection, user grouping design, joint BS's beamforming optimization, and group time allocation. The non-convex expression of the SINR makes the problem more complicated. To effectively solve it, we propose a three-step framework to tackle this cooperative transmission problem in multi-hop multi-RIS communication systems. Specifically, we first analyze the characteristics of the multi-RIS multi-reflection channels, and reformulate the equivalent channel according to the channel composition so that we can obtain the optimal RISs' phase shift design and
	  define corresponding link weight (BS-RIS links, inter-RIS links, RIS-user links) between any two nodes. Then we construct a weighted connection graph and obtain the multi-hop path with the maximum equivalent channel gain for each user. 
	  To avoid the undesired scattering of RISs, we construct a conflict graph, where each vertex in the graph corresponds to one multi-reflection path, and if two transmission paths cannot satisfy the path separation constraint, we add an edge between them. Based on the conflict graph, we can divide users into different activation groups by constructing MISs.   
      Finally, we jointly optimize the BS's beamforming design and the time allocation for each activation group to guarantee the fairness among users. Details are introduced in the next section.

	\section{Framework Design for Cooperative Multi-reflection Transmission \\ in Multi-RIS Wireless Communication System}
	In this section, we introduce the proposed three-step framework for cooperative multi-refection transmission design. In particular, we first find the multi-reflection paths with the maximum equivalent channel gain for each user. Then, we construct the conflict graph to model the interference among the selected paths and divide users into different activation groups. With the obtained activation groups, we propose an iterative optimization algorithm to update the active beamforming at the BS and the group scheduling policy to maximize the minimum data rate of all users. Complexity analysis of the entire system design is also given in this section. 
\vspace{-0.5cm}
\subsection{Multi-reflection Path Selection}
As analyzed before, for a given path $\Omega(k)$, the equivalent channel from the BS to user $k$ is expressed as 
\begin{equation}
\setlength{\abovedisplayskip}{0.8pt}
\setlength{\belowdisplayskip}{0.8pt}
	\begin{aligned} \label{Equ_Channel1}
		&\bm{h}_{0,J+k}\left(\Omega(k)\right)= \\
		&\bm{g}_{a_{N_k}^{(k)},J+k}^H \mathbf{\Phi}_{a_{N_k}^{(k)}}
		\left( \prod_{n \in \mathcal{N}_k,n\neq N_k} \!\!\!\!\!\!\!\! \mathbf{S}_{a_n^{(k)},a_{n+1}^{k}} \mathbf{\Phi}_{a_n^{(k)}}
		\right) \mathbf{H}_{0,a_1^{(k)}}.
	\end{aligned}
\end{equation}
 The channel gain of the equivalent channel $\bm{h}_{0,J+k}\left(\Omega(k)\right)$ depends on the $N_k$ selected RISs and their corresponding phase shifts. To obtain the insights, we reformulate the equivalent channel, by substituting the detailed expression of each individual link, i.e., \eqref{H} - \eqref{g}, into \eqref{Equ_Channel1}, we have 
\begin{equation}
	\begin{aligned} \label{Equ_Channel_re}
		\bm{h}_{0,J+k}\left(\Omega(k)\right)=\frac{\sqrt{\beta_{0}}}{d_{a_{N_k}^{(k)},J+k}} \left(
		\prod_{n=1}^{N_k} A_{n-1,n}^{(k)}
		\right) \tilde{\bm{h}}^H_{0,j,1},
	\end{aligned}
\end{equation}
where
\begin{equation}\label{channel_weight}
	A_{n-1,n}^{(k)}\!\!\!= \!\! 		 \left\{
	\begin{aligned} 	
		&
		\frac{\sqrt{\beta_0}}{d_{0,a_1^{(k)}}}
		\tilde{\bm{s}}^H_{a_1^{(k)},a_2^{(k)},1} \mathbf{\Phi}_{a_1^{(k)}}\tilde{\bm{h}}_{0,a_1^{(k)},2},\\ &\quad\quad\quad\quad\quad\quad\quad\quad\quad\quad\quad\quad if \quad n=1, \\
		&
		\frac{\sqrt{\beta_0}}{d_{a_{n-1}^{(k)},a_n^{(k)}}}
		\tilde{\bm{s}}^H_{a_n^{(k)},a_{n+1}^{(k)},1}\mathbf{\Phi}_{a_n^{(k)}}
		\tilde{\bm{s}}_{a_{n-1}^{(k)},a_{n}^{(k)},2}, \\
		&\quad\quad\quad\quad\quad\quad\quad\quad\quad\quad\quad\quad if \quad 2 \leqslant n \leqslant N_k-1,\\
		&
		\frac{\sqrt{\beta_0}}{d_{a_{N_k-1}^{(k)},a_{N_k}^{(k)}}}
		\tilde{\bm{g}}^H_{N_k,J+k}\mathbf{\Phi}_{a_{N_k}^{(k)}}
		\tilde{\bm{s}}_{a_{N_k-1}^{(k)},a_{N_k}^{(k)},2},\\
		& \quad\quad\quad\quad\quad\quad\quad\quad\quad\quad\quad\quad if \quad  n=N_k.
	\end{aligned} 
	\right.
\end{equation}
For given reflection path $\Omega(k)$, we want to maximize the equivalent channel gain $||\bm{h}_{0,J+k}\left(\Omega(k)\right)||^2$ for optimal received signal strength, thereby the magnitude of each $A_{n-1,n}^{(k)}$ should be optimized. In the following,
we take $2\leqslant n \leqslant N_k-1$ as an example to show the process about how to design RISs' configurations. Specifically, $|A_{n-1,n}^{(k)}|^2$ can be reformulated as
\begin{equation}
	\begin{aligned}\label{Channel_gain}
		|A_{n-1,n}^{(k)}|^2&=\left|\frac{\sqrt{\beta_0}}{d_{a_{n-1}^{(k)},a_n^{(k)}}}\bm{\phi}_{a_n^{(k)}}^H\diag\left(\tilde{\bm{s}}^H_{a_n^{(k)},a_{n+1}^{(k)},1}\right)\tilde{\bm{s}}_{a_{n-1}^{(k)},a_{n}^{(k)},2}\right|^2\\
	&	=\left|\frac{\sqrt{\beta_0}}{d_{a_{n-1}^{(k)},a_n^{(k)}}}\bm{\phi}_{a_n^{(k)}}^H \bm{\rho}_{a_n^{(k)}}\right|^2,
	\end{aligned}
\end{equation} 
where $\bm{\rho}_{a_n^{(k)}}=\diag\left(\tilde{\bm{s}}^H_{a_n^{(k)},a_{n+1}^{(k)},1}\right)\tilde{\bm{s}}_{a_{n-1}^{(k)},a_{n}^{(k)},2}$. 
For continuous RIS beamforming, which means that RIS can achieve any phase shift values in $[0, 2\pi)$, the RIS's configuration is directly optimized according to
\begin{equation}
\setlength{\abovedisplayskip}{2pt}
\setlength{\belowdisplayskip}{2pt}
	\begin{aligned}
		\phi_{a_n^{(k)},m}^{op}=e^{j\angle \left(\bm{\rho}_{a_n^{(k)},m}\right)},\quad m=1,2,...,M_{a_n^{(k)}},
	\end{aligned}
\end{equation} 
and we have $|A_{n-1,n}^{(k)}|^2=\beta_0M^2_{a_n^{(k)}}/d^2_{a_{n-1}^{(k)},a_n^{(k)}}$. For discrete RIS beamforming, which means that the phase shift value of each RIS element can only be selected from a given codebook, we have $\phi_{a_n^{(k)},m}^{op}=\arg \max_{\,\phi_{a_n^{(k)},m}\in \mathcal{B}} 
\left\lbrace \bm{\phi}^H_{a_n^{(k)}} \bm{\rho}_{a_n^{(k)}} \right\rbrace$, where $\mathcal{B}$ represents the given codebook. For both continuous and discrete cases, the optimal value of $|A_{n-1,n}^{(k)}|$ is always proportional to the number of reflection elements at RIS $a_n^{(k)}$, and inversely proportional to the distance between nodes $a_{n-1}^{(k)}$ and $a_n^{(k)}$. Based on this observation, the optimal $|A_{n-1,n}^{(k)}|^2$ can be treated as a constant once the system is deployed, thereby we can use this property to build the connection graph of the whole system. 

From \eqref{Equ_Channel} - \eqref{Channel_gain}, we observe that the individual channel gain from node $a_{n-1}^{(k)}$ to $a_{n}^{(k)}$, $ \forall n \in \mathcal{N}_k \cup \lbrace 0 \rbrace $ can be characterized by $|A_{n-1,n}^{(k)}|^2$, and $|A_{n-1,n}^{(k)}| $ is always proportional to the number of reflection elements at RIS $a_{n}^{(k)}$  and inversely proportional to the transmission distance between node $a_{n-1}^{(k)}$ and $a_{n}^{(k)}$. Based on this, we define the channel weight from node $a_{n-1}^{(k)}$ to $a_{n}^{(k)}$, $ \forall n \in \mathcal{N}_k $ as
	\begin{equation}
		\setlength{\abovedisplayskip}{0.8pt}
		\setlength{\belowdisplayskip}{0.8pt}
		\begin{aligned}
			\digamma_{a_{n-1}^{(k)},a_{n}^{(k)}}=\log\left(1+ \frac{d_{a_{n-1}^{(k)},a_{n}^{(k)}}^2}{\beta_0 M^2_{a_n^{(k)}}}\right),
		\end{aligned}
	\end{equation}
	which is the logarithmic form of the reciprocal of $|A_{n-1,n}^{(k)}|^2$, where $\left(1+ d_{a_{n-1}^{(k)},a_{n}^{(k)}}^2/\beta_0 M^2_{a_n^{(k)}}\right)$ can ensure that the weights of all links are always greater than $0$, thereby  each link will lead to an increasment in the final weights of the overall path, then we can avoid the loop situation as addressed in \eqref{C21}. Similarly, we define the weight of the link from RIS $a_{N_k}^{(k)}$ to user $k$ as
	\begin{equation}
		\begin{aligned}
			\digamma_{a_{N_k}^{(k)},J+k}=\log\left(1+\frac{d^2_{a_{N_k}^{(k)},J+k}}{\beta_0}\right).
		\end{aligned}
	\end{equation}
	Based on the definition of these link weights, we recast the reflection path selection as an equivalent problem in graph theory. We construct a directed and weighted connection graph $G_0=(V_0,E_0)$. The vertex set $V_0$ consists of all nodes in the system, i.e., $V_0 = \lbrace 0, 1, 2,...,J, J+1,..., J + K \rbrace$ and the edge set $E_0$ is defined as 
	\begin{equation}
		\begin{aligned}
			E_0=&\lbrace (0,j)| l(0,j)=1, j\in \mathcal{J}   \rbrace 
			\cup \lbrace (i,j)|l(i,j)=1,i,j\in\mathcal{J}                                                      \rbrace \\
			&\cup
			\lbrace (j,J+k)|l(j,J+k)=1, j\in \mathcal{J}, k\in \mathcal{K}
			\rbrace ,
		\end{aligned}
	\end{equation}
	which means that there exists an edge from vertex $i$ to vertex $j$ if and only
	if they can communicate with each other through LoS link. Given the constructed graph $G_0$, any reflection
	path from the BS to user $k$ corresponds to a path from node $0$ to node $J + k$ in the connection graph $G_0$. Based on this transformation, we adopt the well-known Dijkstra's algorithm to find the multi-reflection paths with the maximum equivalent channel gain for each user, and use $p(k)$, $k \in \mathcal{K}$ to denote the multi-reflection path with the maximum equivalent channel gain from BS to user $k$. In this paper, we only consider the optimal multi-hop transmission path for each user due to the following two considerations:
(1) The overall number of RISs in the system is limited, if we use it to build sub-optimal paths for one user, it will occupy the RISs that can construct the optimal paths for other users. 
(2) The maximal transmit power of the base station is fixed, if sub-optimal paths are considered, it will occupy the power that can be allocated to the optimal path. In other words, if we do not consider the sub-optimal paths, we can concentrate the power on the optimal paths to improve the received signal quality.

	\subsection{ Activation Group Design}	
    To avoid severe inter-path interference caused by undesired scattering of RISs, we introduce the new path separation constraint as shown in \eqref{C31}, only the paths that do not interfere with each other can be activated at the same time for signal transmission. To guarantee this, we need divide users into different activation groups. First, we construct a new undirected conflict graph $G = (V,E)$, where each vertex in $V$ corresponds to one multi-reflection path obtained before, i.e., $V=\left\lbrace p(k)|k\in \mathcal{K} \right\rbrace$. For any two vertices in $G$, we add an edge between them if they cannot satisfy the path separation constraint shown in \eqref{C31}, which means they cannot be activated at the same time. To construct the activation groups, we first introduce the concept of the maximal independent set \cite{MIS}.
	
	\vspace{0.2cm}
	\fbox{ 
		\parbox{0.43\textwidth}{
			\emph{Maximal independent set (MIS): given a conflict graph, an independent set $\mathcal{I}$ is a set of vertices, in which there is no edge between any two of the vertices. This equivalently indicates the paths that can transmit simultaneously in the network. If adding any one more vertex into an independent set $\mathcal{I}$ results in a non-independent set, $\mathcal{I}$ is a maximal independent set.  }
	}}
	\\
	
	Let $\mathcal{I}_1, \mathcal{I}_2,...,\mathcal{I}_Q$ denote the $Q$ MISs found in the conflict graph, and $t_q$, $0 \leq t_q \leq 1$ denote the fraction of time allocated to the MIS $\mathcal{I}_q$, we have the scheduling constraint $\sum_{q=1,2,...,Q} t_q=1$, since only one MIS can be active at a time. During the scheduled period, all the MISs take turns for information transmissions.
	The MIS based design can achieve higher system performance by enabling flexible time resource sharing and effective interference mitigation \cite{FANG12,FANG14,FANG15}. In this work, one MIS corresponds to one activation group in this paper. The activation groups $Q$ needs to satisfies the following two requirements. \textbf{Requirement 1:} These $Q$ activation groups need to contain all the users, so that every user has access to the network, thereby guarantee the feasibility of the final solution. \textbf{Requirement 2:} To reduce the complexity of system design, we want the overall number of activation groups $Q$ to be as small as possible.

    Based on the above requirement, we adopt the heuristic algorithm proposed in \cite{MIS} to judiciously compute a set of MISs covering all the paths for users. 
	The basic idea  of the algorithm is to define a scheduling index (SI) metric to differentiate the interference level of all the paths.  In particular,
	we sort all the vertices $V$ in graph $G$ from low to high according to the number of edges connected to it. The vertex with the minimum number of edges will be assigned with the highest priority. 
	If two vertices have the same number of edges, we randomly choose one of them to assign the higher priority. The output set $\mathcal{P}$ of this sorting process is that all multi-reflection paths are arranged according to SI values in a descending order.
	Finally, we iteratively compute a set of MISs that covers the multi-reflection paths for all users to guarantee the feasibility of the final solution. The algorithm starts with the consideration of all the sorted paths $p(k) \in \mathcal{P} $ as the anchor paths. Each anchor path leads to a round of independent set computing. For example, given anchor path $p(k)$, 
	we will find one IS that covers $p(k)$ and all its non-interference vertices in the conflict graph $G$. The IS starting from $p(k)$ will be stored in the vector $\bm{s}[p(k)]$. 
	
	A to-be-covered set $\mathcal{T}$ is defined as the set containing the paths yet to be covered by certain IS. The set $\mathcal{T}$ is initialized as $\mathcal{T}=\mathcal{P}$, and each time when a new path is covered by a certain IS, it is removed from $\mathcal{T}$. The operations to search for IS will stop when $\mathcal{T}$ becomes empty. In other words, the algorithms will run recursively along the anchor paths until all the paths are covered by a certain IS, which guarantees that the algorithm can generate a feasible solution that covers all users' multi-reflection paths. The detailed processes are shown in \textbf{Algorithm \ref{alg:MIS}}. 
	
	\begin{algorithm}[htbp] 
		\begin{spacing}{1.08}
		\SetAlgoLined
		\KwIn{$\mathcal{P}$ - the set contains all multi-reflection paths $p(k)$, $k \in \mathcal{K}$ in a descending order of scheduling index.}%输入参数
		\KwOut{A set of activation groups $\mathcal{I}_1$, $\mathcal{I}_2$, ..., $\mathcal{I}_Q$.}%输出
		% \KwResult{Write here the result}
		\textbf{initialization}: $\bm{s}[p(k)]=\emptyset $ for all anchor path $p(k)$, to-be-covered set $\mathcal{T}=\mathcal{P}$.\\
		\textbf{Stage one: find independent set (IS)}\\
		\While{$1$ }
		{
			pick the new anchor path $p(k)$ with the highest scheduling index in the to-be-covered set $\mathcal{T}$, add $p(k)$ into $\bm{s}[p(k)]$,
			$\mathcal{T}=\mathcal{T}\setminus \lbrace p(k) \rbrace$;
			%		\If {$p \in \mathcal{T}$}
			%		{
			%			$\mathcal{T}=\mathcal{T}/\lbrace p \rbrace$\;
			%		}
			\\
			\For{$p (k^{'})\in \mathcal{P}, k^{'}\neq k $,}
			{
				\eIf{there is no edge between $p (k^{'})$ and any vertex in $\bm{s}[p(k)]$  }{
					add $p (k^{'})$ into $\bm{s}[p(k)]$\;
					break\;
				}{
					add $p (k^{'})$ into $\bm{s}[p (k^{'})]$;
				}
				\If{$p (k^{'}) \in \mathcal{T}$}
				{
					$\mathcal{T}=\mathcal{T} \setminus\lbrace p (k^{'})\rbrace$\;
				}
			}
			\If{$\mathcal{T} = \emptyset$ (i.e., all paths have been covered)}
			{
				break\;
			}	
		}
		\textbf{Stage two: extend to the maximal independent set (MIS)}\\
		\For{ every independent set found above}
		{
			\For{ $ \forall \, p(k^{'}) \notin  \, \bm{s}[p(k)] $} 
			{	
				check whether there exist edge between $p(k^{'})$ and each vetex in $\bm{s}[p(k)]$;\\
				\If{there is no edge }
				{
					add $p(k^{'})$ to $\bm{s}[p(k)]$ \;
				}	
			}
		}
		\caption{Activation Group Design}
		\label{alg:MIS}
		\end{spacing}
	\end{algorithm}

	\subsection{Joint Optimization of the BS's Beamforming and the Activation Group Scheduling}
	With the obtained activation groups, we can jointly optimize the BS's beamforming and the activation group scheduling to guarantee the fairness of end users. Specifically, with $Q$ obtained activation groups $\mathcal{I}_1, \mathcal{I}_2,...,\mathcal{I}_Q$, the original optimization problem $\textbf{(P1)}$ is reduced to 
		\begin{align}
			\textbf{(P2)} \quad \quad &\max_{t_q,\bm{w}_k^{q}} \, \min_{k \in \mathcal{K}} \quad C_k=\sum_{k \in \mathcal{I}_{q}} t_q\log(1+\gamma_k^{q}), \label{C220}	\\ 
			&\quad \quad \quad s.t.  
			 \sum_{q=1,2,...,Q} \!\! \!\!\! t_q=1, \nonumber \\
			 &\quad\quad \quad\quad \quad \quad    0 \leq t_q \leq 1, \,\, q=1,2,\ldots, Q, \label{C221}\\ 
			&\quad \quad \quad \quad\quad\,\, \sum_{k \in \mathcal{I}_{q}} ||\bm{w}_k^{q}||^2 \leq P_T\label{C222},
		\end{align}
	which is a joint beamforming and group scheduling problem. To solve it, we decouple \textbf{(P2)} into two sub-problems, one for optimizing the beamforming design at the BS and the other to update the group scheduling policy. 
	
	\emph{\textbf{1) Optimize the BS's beamforming.}}	 Specifically, with given group  scheduling parameters $0 \leq t_q \leq 1, q=1,2,\ldots,Q$, $\sum_{q=1}^Q t_q=1$, problem $\textbf{(P2)}$ can be further reduced to $\textbf{(P2.1)}$ as shown in the follow
	\begin{equation}
		\setlength{\abovedisplayskip}{0.8pt}
		\setlength{\belowdisplayskip}{0.8pt}
		\begin{aligned}
			\textbf{(P2.1)} \quad \quad &\max_{\bm{w}_k^{q}} \, \min_{k \in \mathcal{K}} \quad C_k=\sum_{k \in \mathcal{I}_{q}} t_q\log(1+\gamma_k^{q}), \label{C310}	\\ 
			&\quad \quad \quad s.t.  
			\,\,\,\, \sum_{k \in \mathcal{I}_{q}} ||\bm{w}_k^{q}||^2 \leq P_T,
		\end{aligned}
	\end{equation}	
	The decoupled sub-problem $\textbf{(P2.1)}$ is still hard to solve due to the non-convex SINR expression. In order to cope with the complicated objective function in \eqref{C310}, we introduce an auxiliary variable $\Gamma$ and transform $\textbf{(P2.1)}$ to $\textbf{(P2.1.1)}$ as follows
		\begin{align}
		\setlength{\abovedisplayskip}{0.8pt}
		\setlength{\belowdisplayskip}{0.8pt}
			\textbf{(P2.1.1)} \quad \quad &\max_{\bm{w}_k^{q}}  \quad \Gamma, \label{C2110}	\\ 
			&\quad  s.t.  \quad  C_k \geq \Gamma ,\,\,\forall k \in \mathcal{K},\label{C2111} \\
			&\quad\quad\quad \sum_{k \in \mathcal{I}_{q}} ||\bm{w}_k^{q}||^2 \leq P_T\label{C2112},
		\end{align}
	where $C_k=\sum_{k \in \mathcal{I}_{q}} t_q\log(1+\gamma_k^{q})$, $k \in \mathcal{K}$.
	Note that in this new problem $\textbf{(P2.1.1)}$, the objective function \eqref{C2110} and the constraint \eqref{C2112} are all convex, the challenge mainly lies in the constraint \eqref{C2111} due to the non-convexity of the SINR expression and the summation representation of $C_k$. First, to handle the summation representation of $C_k$, we introduce an auxiliary vector $\bm{v}=[v_1,v_2,...,v_K]$, where each parameter $v_k$, $k \in \mathcal{K}$ represents the number of activation groups that user $k$ belongs to. With the auxiliary vector $\bm{v}$, we can further transform $\textbf{(P2.1.1)}$ into
		\begin{align}
			\textbf{(P2.1.2)} \quad \quad &\max_{\bm{w}_k^{q}}  \quad \Gamma, \label{C2120}	\\ 
			&\quad  s.t.  \quad  t_q\log(1+\gamma_k^{q}) \geq \Gamma_k ,\,\,\forall k \in \mathcal{K}, \label{C2121} \\
			&\quad\quad\quad \sum_{k \in \mathcal{I}_{q}} ||\bm{w}_k^{q}||^2 \leq P_T\label{C2122},
		\end{align}
	where $\Gamma_k=\Gamma / v_k$. By doing so, we divide the equivalent data rate of each user to the associated activation groups, and convert the equivalent rate constraint during the whole scheduled time to the rate constraint in each activation group. Another challenge we need to solve is the non-convexity of the SINR expression $\gamma_k^{q}$. To tackle this problem, we further transform \eqref{C2121} as
	\begin{equation}
	\setlength{\abovedisplayskip}{2pt}
	\setlength{\belowdisplayskip}{2pt}
		\begin{aligned} \label{123}
			&t_q\log(1+\gamma_k^{q}) \geq \Gamma_k \\
			&\quad \quad \Leftrightarrow  -\frac{1}{2^{\Gamma_{k,q}}-1}|\bm{h}_{0,J+k}\left(\Omega(k)\right)\bm{w}_k^{q}|^2
			+\\	
			&\quad\quad\quad\quad\sum_{u\in \mathcal{I}_{q},u \neq k}|\bm{h}_{0,J+k}\left(\Omega(k)\right)\bm{w}_u^{q}|^2+\sigma_0^2 \leq 0,
		\end{aligned}
	\end{equation}
	where $\Gamma_{k,q}=\Gamma_{k}/t_q$. This constraint is still hard to solve due to the following two reasons. First, variable $\Gamma_{k,q}$ appears in the denominator. Second, the term $-\frac{1}{2^{\Gamma_{k,q}}-1}$ is negative when $\Gamma_{k,q} >0$. Therefore, this constraint is still non-convex.
	
	To deal with the variable $\Gamma_{k,q}$ in the denominator, we construct a bisection framework to first set $\Gamma$ to a certain value $\Gamma_0$ in advance, and then solve the corresponding feasibility problem
	to examine whether the minimum rate can achieve $\Gamma_0$. In this way, the constraint becomes convex within $\Gamma_0$. To be more specific, we utilize the total consumed power as the indicator for feasibility, and minimize this power while ensuring that all users' equivalent rates are larger than $\Gamma_0$. If the minimized power consumption is smaller than the maximum transmission power $P_T$, $\Gamma_0$ can be achieved, and otherwise $\Gamma_0$ cannot be achieved \cite{Dai1}. The reformulated problem for each activation group is expressed as 
		\begin{align}
		\setlength{\abovedisplayskip}{2pt}
		\setlength{\belowdisplayskip}{2pt}
			\textbf{(P2.1.3)} \,\,  &\min_{\bm{w}_k^{q}}  \quad \sum_{k \in \mathcal{I}_{q}} ||\bm{w}_k^{q}||^2 ,\label{C2130}	\\ 
			&  s.t.  \quad -\frac{1}{2^{\Gamma_{0,k,q}}-1}|\bm{h}_{0,J+k}\left(\Omega(k)\right)\bm{w}_k^{q}|^2 \nonumber\\
			& \quad\quad\quad+	\sum_{u\in \mathcal{I}_{q},u \neq k}|\bm{h}_{0,J+k}\left(\Omega(k)\right)\bm{w}_u^{q}|^2+\sigma_0^2 \leq 0, \label{NewC}
		\end{align}
	where $\Gamma_{0,k,q}=\Gamma_{0}/(v_k t_q)$. Once we figure out that $\Gamma_{0}$ is achievable or not by solving a series of problems $\textbf{(P2.1.3)}$, a bisection procedure can be carried out to acquire
	the maximum $\Gamma$. Problem $\textbf{(P2.1.3)}$ is still
	hard to solve, because the new constraint \eqref{NewC} is non-convex due to the negativity of the term $-\frac{1}{2^{\Gamma_{0,k,q}}-1}$. To address this issue, we further rewrite the mathematical expressions of the beamforming vectors $\bm{w}_k^{q}$ into the form of matrix trace as
	\begin{equation}
	\setlength{\abovedisplayskip}{2pt}
	\setlength{\belowdisplayskip}{2pt}
		\begin{aligned}
			||\bm{w}_k^{q}||^2=\text{Tr}\left(\bm{w}_k^{q}{(\bm{w}_{k}^{q})}^H\right) =\text{Tr}\left( \mathbf{W}_k^{q}\right),
		\end{aligned}
	\end{equation}
	\begin{equation}
	\setlength{\abovedisplayskip}{2pt}
	\setlength{\belowdisplayskip}{2pt}
		\begin{aligned}
			&|\bm{h}_{0,J+k}\left(\Omega(k)\right)\bm{w}_k^{q}|^2 \\&=\text{Tr}
			\left(
			\bm{h}_{0,J+k}\left(\Omega(k)\right){\bm{h}^H_{0,J+k}\left(\Omega(k)\right)\bm{w}_k^{q}(\bm{w}_k^{q}})^H 
			\right)\\
			&=\text{Tr}\left(
			\bm{h}_{0,J+k}\left(\Omega(k)\right)\bm{h}^H_{0,J+k}\left(\Omega(k)\right)\mathbf{W}_k^{q}
			\right),
		\end{aligned}
	\end{equation}
	where $\bm{w}_k^{q}$ is lifted to $ \mathbf{W}_k^{q}=\bm{w}_k^{q}{(\bm{w}_{k}^{q})}^H$. Then the problem $\textbf{(P2.1.3)}$ is equivalently rewritten as
		\begin{align}
			\textbf{(P2.1.4)} &\min_{\bm{w}_k^{q}}  \sum_{k \in \mathcal{I}_q} \, \text{Tr}\left( \mathbf{W}_k^{q}\right) ,\label{C2140}	\\ 
			& s.t.   -\frac{1}{2^{\Gamma_{0,k,q}}\!-\!1}\text{Tr}\left(
			\bm{h}_{0,J+k}\left(\Omega(k)\right)\bm{h}^H_{0,J+k}\left(\Omega(k)\right)\mathbf{W}_k^{q}
			\right) \nonumber\\
			&\quad\quad\quad\quad	+\!\!\!\!\!\!\!\!	\sum_{u\in \mathcal{I}_{q},u \neq k}\!\!\!\!\!\!\text{Tr}\left(
			\bm{h}_{0,J+k}\left(\Omega(k)\right)\bm{h}^H_{0,J+k}\left(\Omega(k)\right)\mathbf{W}_u^{q}
			\right) \nonumber\\
			&\quad\quad\quad\quad+\sigma_0^2 \leq 0, \label{NewC1}\\
			& \quad\quad\quad \quad \mathbf{W}_u^{q} \in \mathcal{H}^{+}_{M_0}, u \in \mathcal{I}_q ,
		\end{align}
	where $\mathcal{H}^{+}_{M_0}$ denotes the set for all Hermitian positive semidefinite matrices of dimension $M_0 \times M_0$. We can find that now constraint \eqref{NewC1} becomes a convex constraint, since $ \mathbf{W}_k^{q}=\bm{w}_k^{q}{(\bm{w}_{k}^{q})}^H$ and $\mathbf{W}_k^{q}$ must be a positive semi-definite matrix. Thus, the problem $\textbf{(P2.1.4)}$ becomes convex, and is ready to be solved for examination $\Gamma_0$. Denote the solution to the problem and the optimal value of the objective function as $\Gamma^{*}$ and $P^{*}$, we have  $\Gamma^{*}\geq \Gamma_0$ when $P^{*} \leq P_T$, and $\Gamma^{*}\leq \Gamma_0$ when $P^{*} \geq P_T$. The proposed bisection-based beamforming method is outlined in \textbf{stage one} of \textbf{Algorithm 2}. Finally, the eigenvalue decomposition is carried out for $\left(\mathbf{W}_k^{q}\right)^{*}$ and the optimal $(\bm{w}_k^{q})^{*}$ is set to be the eigenvector with the maximum eigenvalue.
	
	\emph{\textbf{2) Update the activation group scheduling policy.}} With the updated beamforming vectors, the sub-problem for optimizing user scheduling can be expressed as
		\begin{align}
			\textbf{(P2.2)} \quad \quad &\max_{t_q} \, \min_{k \in \mathcal{K}} \quad C_k=\sum_{k \in \mathcal{I}_{q}} t_q\log(1+\gamma_k^{q}), \label{C320}	\\ 
			&\quad \quad \quad s.t.  
			\,\, \sum_{q=1,2,...,Q} t_q=1, \nonumber\\
			&\quad\quad\quad\quad\quad\quad\quad\quad 0 \leq t_q \leq 1, \,\, q=1,2,\ldots, Q
			\label{C321}.
		\end{align}
	Note that with given beamforming vectors $\bm{w}_k^{q}$, $\gamma_k^{q}$ becomes a constant in the objective function \eqref{C320}.
	% Then, we introduce an auxiliary variable  $\Upsilon$ and transform $\textbf{(P2.2)}$ into the following problem
	%\begin{subequations}
	%	\begin{align}
	%		\textbf{(P2.2.1)} \quad \quad &\max_{t_i} \quad \quad \Upsilon
	%		\label{C330}	\\ 
	%		&\quad  s.t.  
	%		\,\, \sum_{i=1,2,...,I}t_i=1\label{C331},\\
	%	&\quad\quad\quad\quad	\sum_{k \in \mathcal{I}_{i}} t_i\log(1+\gamma_k^{t_i}) \geq \Upsilon \label{NEWC2}.
	%	\end{align}
	%\end{subequations}
	%Note that with given beamforming vectors $\bm{w}_k^{t_i}$,  $\gamma_k^{t_i}$ in \eqref{NEWC2} becomes a constant value,
	As a result, in  problem $\textbf{(P2.2)}$, both the objective function and the constraints are linear. $\textbf{(P2.2)}$ can be directly solved by the classical convex optimization methods, for example, linear programming. The detailed processes for updating the scheduling parameters are summarized in \textbf{stage two} of \textbf{Algorithm 2}.

		\begin{algorithm}[htb] 
			\begin{spacing}{1.08}
		\SetAlgoLined
		\KwIn{The activation groups $\mathcal{I}_1$, $\mathcal{I}_2$, ..., $\mathcal{I}_Q$; the multi-reflection pathes $\Omega(k)$ for each user; maximum transmit power $P_T$ at the BS; noise power $\sigma^2_0$; desirable accuracy $\varepsilon$ and $\epsilon$, the maximum number of iterations $N$.}%输入参数
		\KwOut{The equvalient data rate $C_k$, $k=1,2,...,K$, for each user; beamforming vector $\bm{w}_k^{q}$, $k=1,2,...,K$, $q=1,2,...,Q$; scheduling pamameter $t_q$, $q=1,2,...,Q$.}%输出
		% \KwResult{Write here the result}
		\textbf{initialization}: $\Gamma_L=0$, $\Gamma_H=\log(1+P_T ||\bm{h}^{max}_{0,J+k}\left(\Omega(k)\right)||^2)/\sigma^2$, where $\bm{h}^{\max}_{0,J+k}\left(\Omega(k)\right)$ is the maximum equivalent channel gain among all users; $t_q$ that satisfies $0 \leq t_q \leq 1$, $\sum_{q=1,2,...,Q}t_q=1$.
		\\
		\While {$\min (C_k)^{n}-\min (C_k)^{n-1} \geq \varepsilon$ or the number of iteration times $n \leq N$}
		{
			\textbf{Stage one: beamforming optimization at the BS}
			
			\While{$\Gamma_H-\Gamma_L \geq \epsilon$ }
			{
				Set $\Gamma_0=(\Gamma_H+\Gamma_L)/2$, solve the optimization
				problem $\textbf{(P2.1.4)}$ to obtain the matries $\mathbf{W}_k^{q}$;
				\\
				\eIf{ $\sum_{k \in \mathcal{I}_q} \text{Tr}\left( \mathbf{W}_k^{q}\right) \leq P_T$, $\forall q=1,2,...,Q$ }{
					Set $\left(\mathbf{W}_k^{q}\right)^{*}=\mathbf{W}_k^{q}$, $\Gamma_L=\Gamma_0$ \;
				}{ Set $\Gamma_H=\Gamma_0$ \; 
				}
			}
			$\Gamma^{*}=\Gamma_l$;
			\\
			Set $(\bm{w}_k^{q})^{*}$ as the eigenvector of $\left(\mathbf{W}_k^{q}\right)^{*}$ with the maximum eigenvalue.
			\\
			\textbf{Stage Two: user scheduling optimization}\\
			Define $0 \leq t_q \leq 1$, $q=1,2,...,Q$ as the optimization variables;\\
			Update the equivalent rate $C_k$ for each user with the optimized beamforming vectors, which satisftes $\sum_{q=1,2,...,Q}t_q=1$ ;\\
			Solve the linear problem \textbf{(P2.2)} to maximize the minimum equivalent rate;
		}
		\caption{ Joint Beamforming and Scheduling }
		\label{alg:Joint_OPT}
		\end{spacing}
	\end{algorithm} 
\vspace{-0.5cm}
	\subsection{Complexity Analysis of the Overall System Design}
	As discussed above, the system design begins with finding the optimal multi-reflection path for each user.
	The complexity of finding all the paths with the maximum equivalent channel gain for each user is $\mathcal{O}\left(KJ^2\right)$ \cite{COMPLEXITY}.
	With the obtained optimal multi-reflection paths, we are able to divide them into different activation groups as shown in \textbf{Algorithm 1}. The input of \textbf{Algorithm 1} is the set of all paths sorted by the scheduling index, where the complexity of this sorting process has linear complexity $\mathcal{O}(K)$. In each iteration of \textbf{Algorithm 1}, given an anchor path $p(k)$, the number of non-neighbor paths in $\mathcal{I}_n\left(p(k)\right)$ is upper bounded by $K-1$. To determine the number of iterations incurred during the activation group search, we consider the worst case that all paths interfere with each other, in which each path incurs a new independent set. Based on this argument, the number of basic operations determining the activation groups is $\mathcal{O}\left((K^2-K)/2\right)$. After obtaining all the candidate ISs, we should further check the maximality of each set and make necessary extension to MISs. The number of candidate ISs is upper bounded by $K$, and each candidate IS needs at most $K^2$ operations for the extension. Combining the above two processes, the\textbf{ worst case complexity} for \textbf{Algorithm 1} is $\mathcal{O}\left(K^3\right)$. Finally, with the obtained $Q$ activation groups, we can conduct the joint optimization of the BS's beamforming and the group scheduling as shown in  \textbf{Algorithm 2}. The complexity of \textbf{Algorithm 2} mainly comes from the process for beamforming design. Given a solution accuracy $\epsilon$, the complexity for updating the beamforming vectors is given by $\mathcal{O}\left(Q(M_0^4 K^{1/2} \log (1/\epsilon))\right)$ \cite{Com_SDR}, where $Q$ represents the number of activation groups. For updating user scheduling parameters, since the objective function is a linear problem, so it can be directly solved by the classical convex optimization technologies with linear complexity $\mathcal{O}(Q)$. 
	Combining all the above processes, we have the overall complexity  $\mathcal{O}\left(KJ^2+K^3+Q\left(M_0^4 K^{1/2} \log (1/\epsilon)+1\right)T \right)$, where $T$ represents the number of iterations done until \textbf{Algorithm 2} converges.

\section{Performance Evaluation}
In this section, we provide numerical results to evaluate the performance of the proposed multi-reflection transmission framework.
We first investigate the performance of each step in the proposed three-step framework to verify the fairness of the final design, then we simulate the effect of system settings on the minimum equivalent rate.
In our simulations, we consider a square area with side length equals to 20 $\mathrm{m}$. And there are $J=16$ RISs and $K=14$ users.
The 3D coordinates of all nodes, their available LoS communication links, and the facing directions of all RISs are shown in Fig. \ref{fig:Simulation_model}, respectively. The system is assumed to operate at the carrier frequency 5 GHz. Thus, the corresponding wavelength is $\lambda = 0.06 \, \mathrm{m}$ and the LoS path gain at the reference distance $1 \, \mathrm{m}$ is assumed to be calculated as $\beta_0=(\lambda/4\pi)^2=-46.4$ dB \cite{Mei1}. The antenna and element spacing at the BS and each IRS are set to $d_A=d_I=\lambda/2$, respectively. We assume that each RIS has the same number of reflecting elements, i.e., $M_j=20, j=1,2,...,16$, and the BS is equipped with $M_0=20$ antennas.
	
	\subsection{Fairness of the Proposed Design}
	In this section, we first show the discovered multi-reflection paths with maximal equivalent channel gain for each user, then we show the obtained activation groups according to our search. Based on the acquired multi-reflection paths and activation groups, we verify the fairness of our proposed multi-RIS multi-hop transmission design.

	\begin{figure}[t]
	\vspace{-0.5cm}
	\centering
\includegraphics[width=0.45\textwidth]{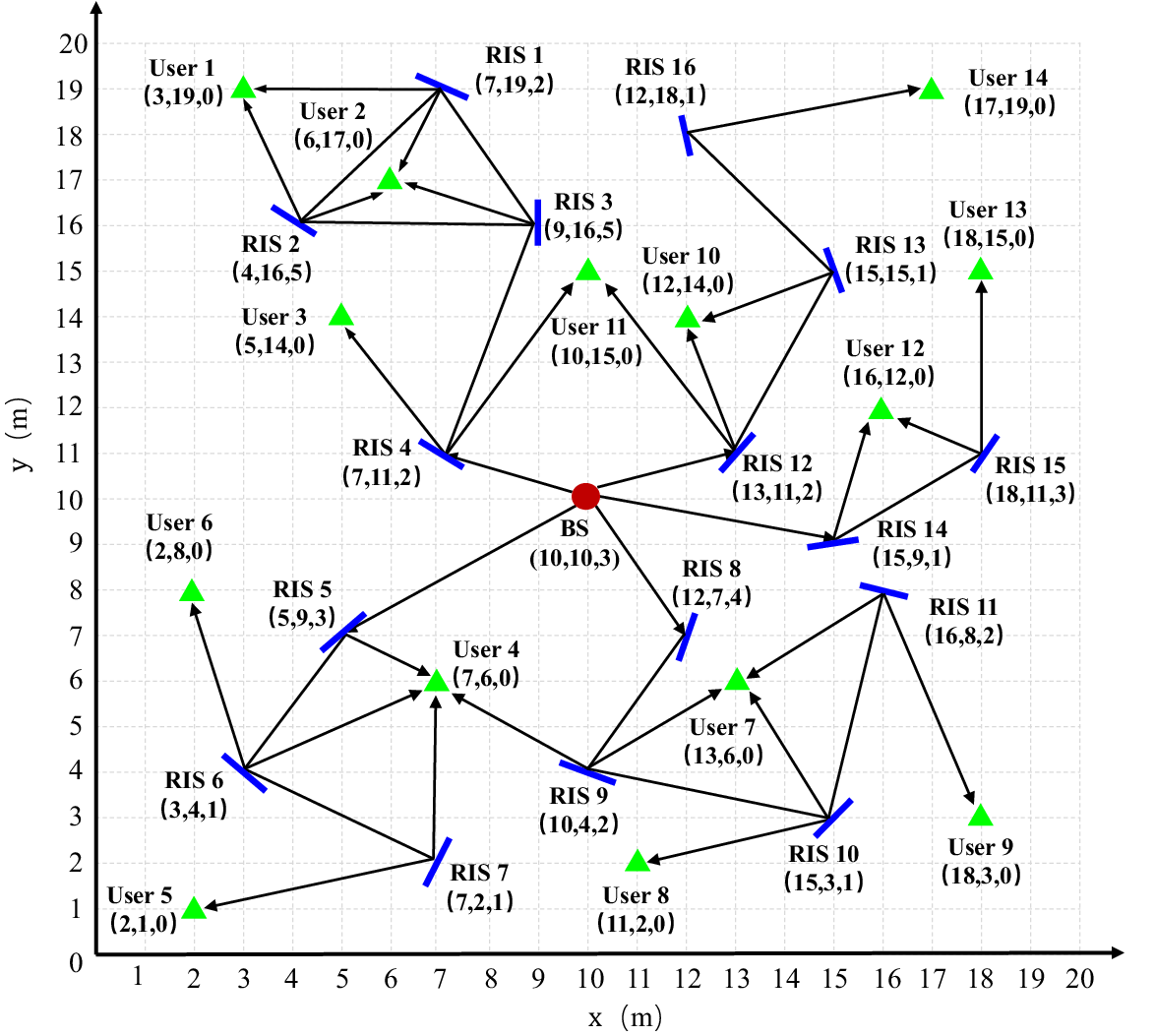}
\caption{Graph representation of the simulation setup.}
\label{fig:Simulation_model}
\end{figure}

	\begin{figure}[t]
	\centering
\includegraphics[width=0.45\textwidth]{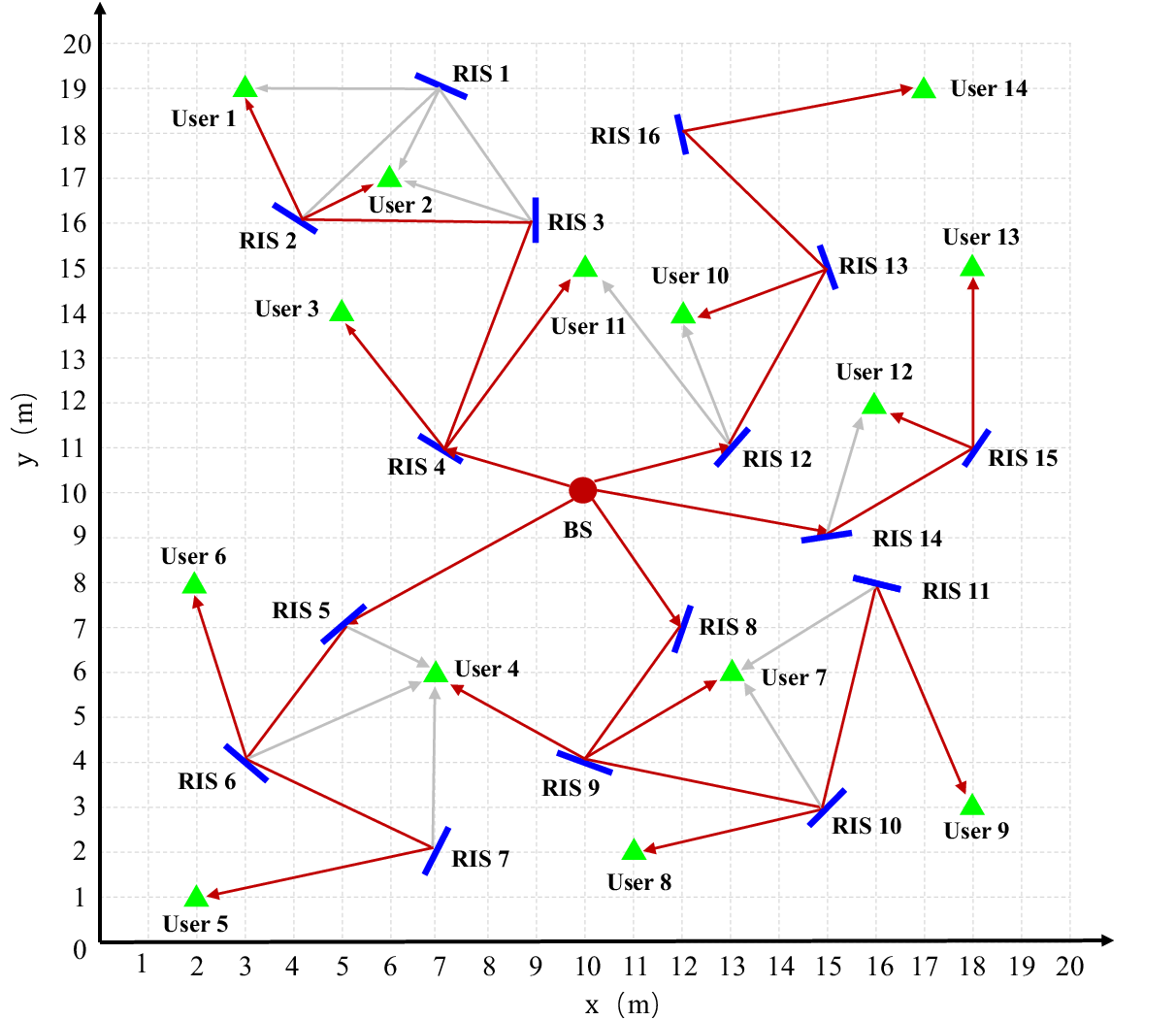}\\
\caption{Designed multi-reflection paths for each user.}
\label{fig:Selected_channels}
\end{figure}

%\begin{figure}[t!]
%	\centering
%	\setlength{\belowcaptionskip}{-0.6cm}
%	\begin{minipage}{0.49\linewidth}
%		\begin{centering}
%			\includegraphics[width=1.02\textwidth]{MXY-Figs/Simulation_model.eps}
%			\caption{Graph representation of the simulation setup.}
%			\label{fig:Simulation_model}
%		\end{centering}
%	\end{minipage}
%	%\qquad
%	\begin{minipage}{0.49\linewidth}
%		\begin{centering}
%			\includegraphics[width=1.06\textwidth]{MXY-Figs/Selected_channels.eps}\\
%			\caption{Designed multi-reflection paths for each user.}
%			\label{fig:Selected_channels}
%		\end{centering}
%	\end{minipage}
%\end{figure}
	
	Fig. \ref{fig:Selected_channels} shows the found multi-reflection paths for all users. By utilizing the Dijkstra's algorithm for the shortest path discovery on graph $G_0$, we plot the optimal multi-reflection paths with the maximum channel gain for each user. We observe from Fig. \ref{fig:Selected_channels} that due to the large number of users and their random locations, there exists conflicts between users' multi-reflection paths. For example, the optimal multi-reflection paths for user $4$ and user $6$ are BS - RIS $8$ - RIS $9$ - user $4$ and BS - RIS $5$ - RIS $6$ - user $6$, respectively. If the BS transmits signals to user $4$ and user $6$ at the same time slot, in addition to the found LoS multi-reflection transmission paths, user $4$ may also receive interference caused by the undesired scattering of RIS $5$ and RIS $6$, since user $4$ is also within the signal interference range of RIS $5$ and RIS $6$. 
	Therefore, it is better to divide users $4$ and $6$ into different activation groups, so that the scattered interference caused by RISs can be well eliminated.
	
	In Table \ref{table_MIS}, we show the obtained activation groups according to our search. Since we assign the scheduling index to each multi-hop path based on the interference level, the path with less interference is assigned at higher priority. In each round of independent set construction, we always start with the paths with higher priority, thereby we can cover all the paths with the minimum number of activation groups. From Table \ref{table_MIS}, we see that we only need four activation groups to cover all the multi-hop transmission paths for all users, which greatly reduces the complexity of subsequent system design. Furthermore, benefiting from the extension operation in stage two of \textbf{Algorithm 1}, we observe that some users can be divided into different activation groups, for example, users $12$ is included in activation groups $1$, $3$ and $4$, user $10$ is included in activation groups $1$ and $3$. As a result, the multi-reflection transmission paths for users $10$ and $12$ can be activated in different time slot, enabling more efficient time resource utilization, yielding higher system performance. 
	
	\renewcommand\arraystretch{1.0}
	\begin{table}[tb]
		\centering
		\caption{\textsc{The Obtained Activation Groups According to Our Policy}}
		\label{table_MIS}
		%	\begin{threeparttable}
		\begin{tabular}{p{4cm}<{\centering}|p{4cm}<{\centering}}
			\hline 
			\textbf{\, }  & The indexes of users  \\
			\hline
		    Activation group 1 &  $3, 6, 8, 10, 12$\\
			\hline
			Activation group 2  & $2, 5, 7, 13, 14$	\\
			\hline 
			Activation group 3 & $1, 4, 10, 12$\\
			\hline	
			Activation group 4 & $5, 9, 11, 12$\\
			\hline
		\end{tabular}	
		%	\end{threeparttable}
	\end{table}

\begin{figure}[t]
	\centering
	\includegraphics[width=0.45\textwidth]{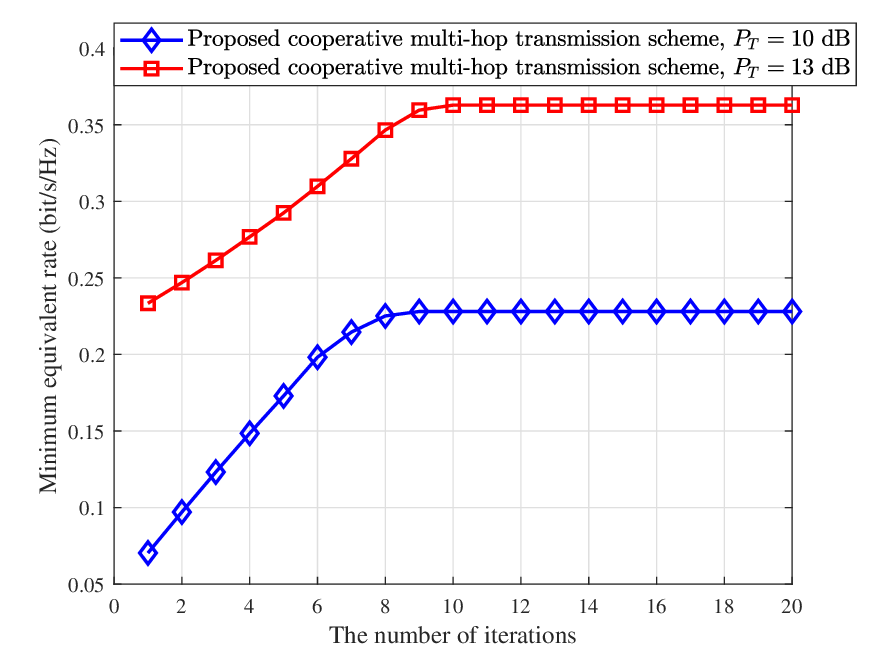}\\
	\caption{The convergence property of the proposed cooperative multi-hop transmission algorithm. The number of antennas at the BS is $M_0=20$. There are $16$ RISs in the system and the number of reflecting elements at each RIS is $M_j=20, j=1,2,...,16$.}
	\label{fig:convergence}
\end{figure}

\begin{figure}[t]
	\vspace{-0.5cm}
	\centering
	\includegraphics[width=0.45\textwidth]{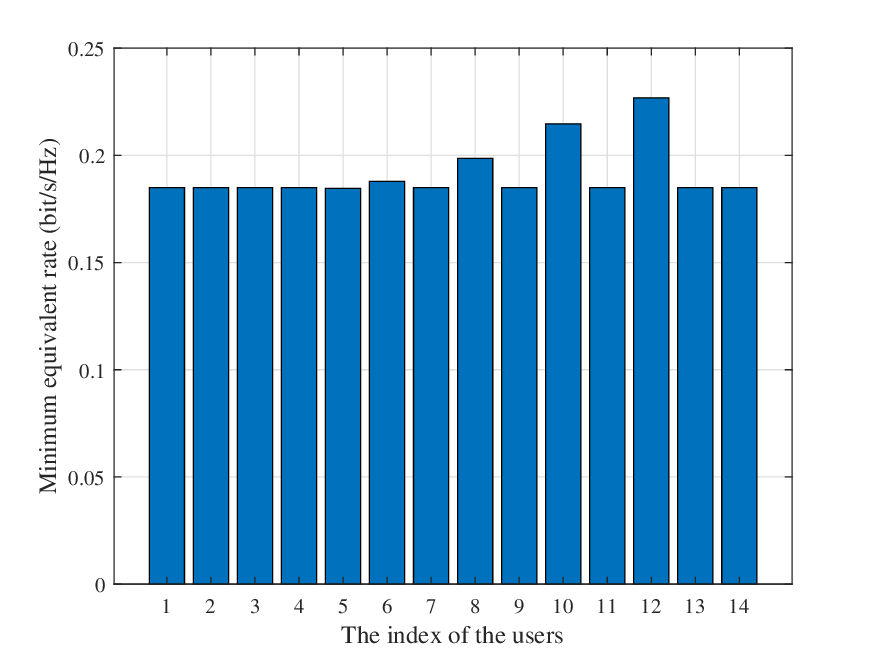}\\
\caption{The equivalent data rate of all users achieved by the proposed framework. The number of antennas at the BS is $M_0=20$. There are $16$ RISs in the system and the number of reflecting elements at each RIS is $20$. The transmit power at the BS is $P_T=10$ dB.}
\label{fig:all_rate}
\end{figure}

%\begin{figure}[tbp]
%	\centering
%	\setlength{\belowcaptionskip}{-1.0cm}
%	\begin{minipage}{0.49\linewidth}
%		\centering
%\includegraphics[width=1.03\textwidth]{MXY-Figs/convergence.eps}\\
%\caption{The convergence property of the proposed cooperative multi-hop transmission algorithm. The number of antennas at the BS is $M_0=20$. There are $16$ RISs in the system and the number of reflecting elements at each RIS is $M_j=20, j=1,2,...,16$.}
%\label{fig:convergence}
%	\end{minipage}
%	%\qquad
%	\begin{minipage}{0.49\linewidth}
%		\centering
%	\includegraphics[width=1.05\textwidth]{MXY-Figs/all_rate.eps}\\
%\caption{The equivalent data rate of all users achieved by the proposed framework. The number of antennas at the BS is $M_0=20$. There are $16$ RISs in the system and the number of reflecting elements at each RIS is $20$. The transmit power at the BS is $P_T=10$ dB.}
%\label{fig:all_rate}
%	\end{minipage}
%\end{figure}

%	\begin{figure}[t]
%	\vspace{-0.5cm}
%	\centering
%	\includegraphics[width=0.45\textwidth]{MXY-Figs/convergence.eps}\\
%	\caption{The convergence property of the proposed cooperative multi-hop transmission algorithm. The number of antennas at the BS is $M_0=20$. There are $16$ RISs in the system and the number of reflecting elements at each RIS is $M_j=20, j=1,2,...,16$.}
%	\label{fig:convergence}
%\end{figure}
	
	Based on the designed activation groups, we first investigate the convergence property of the proposed algorithm for cooperative multi-reflection design.
	In Fig. \ref{fig:convergence}, we show the convergence property and the achievable minimum equivalent rate of \textbf {Algorithm 2} when the transmit power budget at the BS varies. It is observed from Fig. \ref{fig:convergence} that the proposed joint beamforming and user scheduling algorithm converges within several iterations for both cases $P_T= 10$ dBW and $P_T=13$ dBW, respectively. Besides, we observe that higher transmit power at the BS can achieve better minimum equivalent data rate.
	
	To verify the fairness of the proposed design for multi-hop multi-RIS transmission systems, we show the final equivalent data rate of all users in Fig. \ref{fig:all_rate}. We observe from Fig. \ref{fig:all_rate} that except for users $10$ and $12$, every user can achieve almost the same equivalent rate through our transmission framework, which guarantees the fairness among users. For users $10$ and $12$, we observe from the graph representation of the system, i.e., Fig. \ref{fig:Simulation_model}, that since they have less interference with others users, thereby they can be included in different activation groups, i.e., user $10$ is included in activation groups $1$ and $3$, user $12$ is included in activation groups $1$, $3$ and $4$,  thereby they can receive signals in different time slots, leading to higher equivalent date rate compared with other users that can only receive signals in one time slot. This phenomenon further demonstrates the superiority of the proposed transmission framework. Due to the extension operation considered at stage two of \textbf{Algorithm 1}, users with less inter-path interference can be activated in different time slots, leading to more flexible and efficient time resource allocation, hence higher equivalent data rate.
	
%	\begin{figure}[t]
%	\vspace{-0.5cm}
%	\centering
%	\includegraphics[width=0.45\textwidth]{MXY-Figs/all_rate.eps}\\
%	\caption{The equivalent data rate of all users achieved by the proposed transmission framework. The number of antennas at the BS is $M_0=20$. There are $16$ RISs in the system and the number of reflecting elements at each RIS is $20$. The transmit power at the BS is $P_T=10$ dB.}
%	\label{fig:all_rate}
%\end{figure}
\vspace{-0.5cm}
\subsection{Impacts of the System Settings on the Minimum Equivalent Rate}

  In this subsection, we investigate the impacts of the system settings on the minimum equivalent rate. To verify the superiority of the proposed transmission framework for multi-hop multi-RIS transmission systems, we compare it with the following three benchmarks. And to create a fair comparison scenario, in the simulation, we assume that all the benchmark schemes can serve users through LoS paths.
	\begin{itemize}	
		\item 
		\emph{\textbf{Single-reflection scheme.}} Each user can only be served by one RIS that provides the maximum equivalent channel gain. If there exists RIS-scattered interference between any two users, we assign them into different activation groups and perform group scheduling accordingly.  
		
		\item 
		\emph{\textbf{Non-RIS scheme.}} The BS directly transmits signals to users without the help of any RIS. Since no RIS is involved in the signal transmission, the inter-path interference  caused by undesired scattering of RISs does not exist, thereby all users can communicate during the whole scheduling period.	
		
		\item
		\emph{\textbf{Non-interference management scheme \cite{Mei1}.}} Maximum-ratio transmission (MRT) beamforming scheme is adopted at the base station side. For fair comparison, the activation group proposed in our paper is adopted and no further interference mitigation schemes are utilized following \cite{Mei1}.	
	\end{itemize}

\begin{figure}[t]
	\centering
	\includegraphics[width=0.45\textwidth]{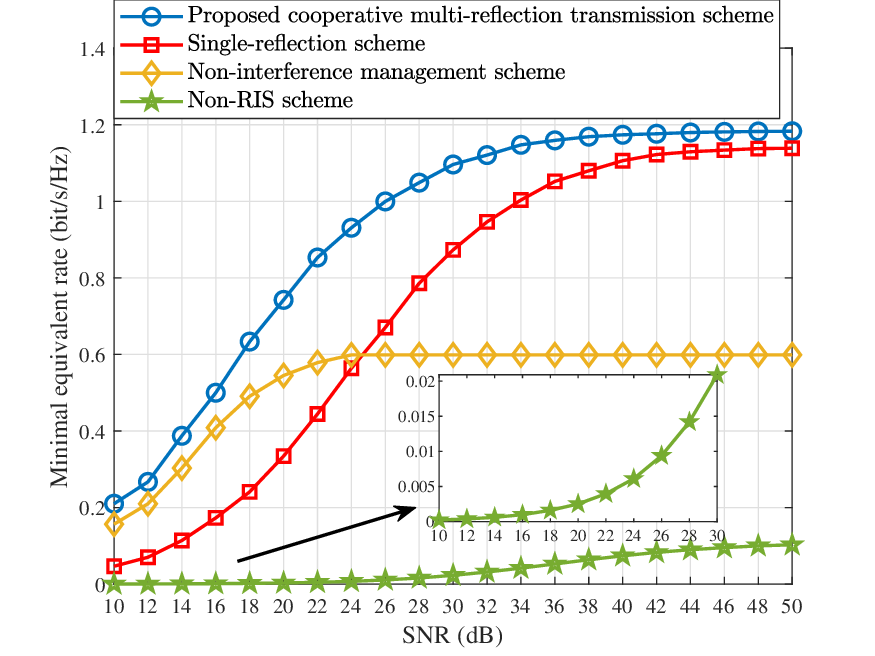}\\
	\caption{The minimum equivalent data rate for different SNR. The number of antennas at the BS is $M_0=20$. There are $16$ RISs in the system and the number of reflecting elements at each RIS is $M_j=20, j=1,2,...,16$.}
	\label{fig:SNR_increasing}
\end{figure}

\begin{figure}[t]
	\centering
	\includegraphics[width=0.45\textwidth]{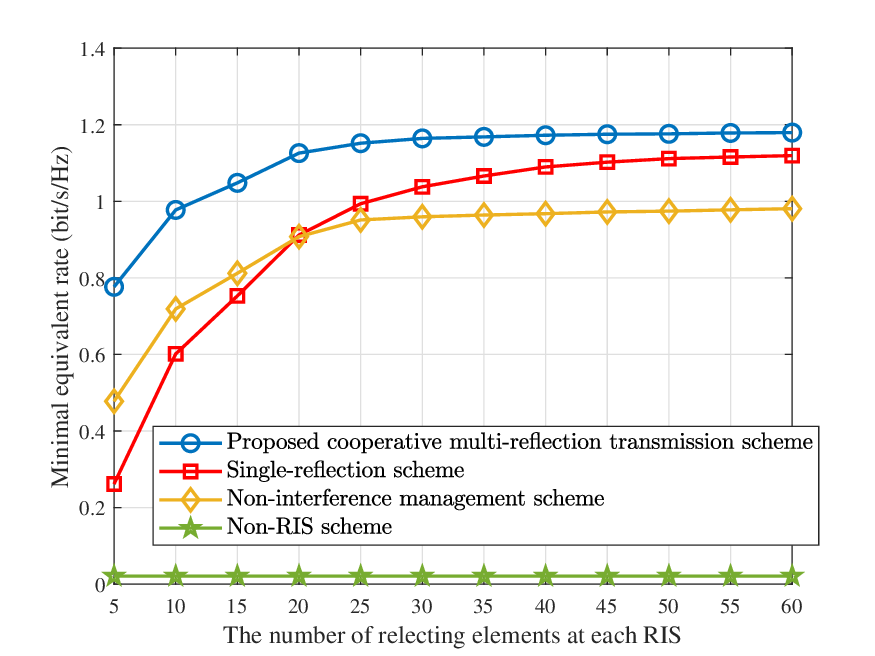}\\
	\caption{The minimum equivalent data rate for different numbers of reflecting elements at each RIS. The number of antennas at the BS is $M_0=20$. There are $16$ RISs in the system, and the  transmit power at the BS is $P_T=30$ dBW. }
	\label{fig:M_increasing}
\end{figure}

%\begin{figure}[tbp]
%	\centering
%	\setlength{\belowcaptionskip}{-1.0cm}
%	\begin{minipage}{0.49\linewidth}
%		\centering
%		\includegraphics[width=1.03\textwidth]{MXY-Figs/SNR_increasing_COM.eps}\\
%		\caption{The minimum equivalent data rate for different SNR. The number of antennas at the BS is $M_0=20$. There are $16$ RISs in the system and the number of reflecting elements at each RIS is $M_j=20, j=1,2,...,16$.}
%		\label{fig:SNR_increasing}
%	\end{minipage}
%	%\qquad
%	\begin{minipage}{0.49\linewidth}
%		\centering
%			\includegraphics[width=1.05\textwidth]{MXY-Figs/M_increasing_COM.eps}\\
%		\caption{The minimum equivalent data rate for different numbers of reflecting elements at each RIS. The number of antennas at the BS is $M_0=20$. There are $16$ RISs in the system, and the  transmit power at the BS is $P_T=30$ dBW. }
%		\label{fig:M_increasing}
%	\end{minipage}
%\end{figure}

%\begin{figure}[t]
%		\vspace{-0.3cm}
%		\centering
%		\includegraphics[width=0.45\textwidth]{MXY-Figs/SNR_increasing.eps}\\
%		\caption{Comparison of the minimum equivalent data rate when SNR increasing. The number of antennas at the BS is $M_0=20$. There are $16$ RISs in the system and the number of reflecting elements at each RIS is $M_j=20, j=1,2,...,16$.}
%		\label{fig:SNR_increasing}
%\end{figure}
	
Fig. \ref{fig:SNR_increasing} shows the minimum equivalent rates of all users when adopting the proposed multi-reflection scheme, single-RIS scheme, non-RIS scheme and non-interference management scheme \cite{Mei1}, respectively. It is seen from Fig. \ref{fig:SNR_increasing} that as the SNR increases, the minimum equivalent rates of all schemes increase, and our proposed cooperative multi-reflection scheme always achieve the best performance. This is because the proposed scheme makes full use of the inter-RIS transmissions to form multi-hop multi-reflection paths, which significantly increases the passive beamforming gain. Besides, the users are divided into different activation groups to avoid the inter-path interference. 
And the carefully designed beamforming pattern and  group scheduling policy further improve system performance. Another noteworthy observation is that when the SNR is very high, for example, SNR= $50$ dB as shown in the figure, the gap between the proposed scheme and single-RIS scheme becomes smaller. The reason is that when the transmit power is extremely high, the inter-user interference becomes the dominating factor that limits the system performance. Since the inter-user interference received at one particular user is related to the equivalent channel gain, normally, the larger the equivalent channel gain, the higher the inter-user interference. In our proposed multi-reflection scheme, the equivalent channel gain for each user is usually larger than single-RIS scenario. Therefore, when SNR is extremely high, the inter-user interference is severer, which limits the improvement of the system performance.
Further, We can observe from Fig. \ref{fig:SNR_increasing} that the non-interference management scheme \cite{Mei1} reaches the bottleneck quickly, since no interference mitigation schemes are considered, which severely limit the overall resource utilization, and the final performance is even worse than single-reflection scheme.

Fig. \ref{fig:M_increasing} shows the minimum equivalent rate of all users obtained from the proposed scheme, single-RIS scheme, non-RIS scheme and non-interference management scheme \cite{Mei1}, respectively. It is shown in  Fig. \ref{fig:M_increasing} that due to the larger equivalent channel gain brought by the multi-reflection paths and the well mitigated interference, our proposed cooperative multi-reflection scheme outperforms the other benchmarks. Among all the comparison schemes, non-RIS scenario has the worst performance due to the non-customized radio transmission environments. 
	The non-interference management scheme \cite{Mei1} outperforms the single-reflection scheme when the number of reflecting elements is small, but the situation reverses with the continuously increase of reflecting elements. This is because with more reflecting elements, the enhanced equivalent channel gain makes the interference more and more severe, which becomes the major factor limiting the furthre growth of system performance. Besides, we find from Fig. \ref{fig:M_increasing} that
there is a diminishing return as the number of reflecting elements increases due to the inter-user interference. Also, the performance gap between the proposed multi-reflection transmission and single-reflection scheme becomes smaller when the number of reflecting elements at each RIS is large. That is because with more reflecting elements, 
the equivalent channel gain of the proposed scheme is much larger than single-RIS scenario, which not only improves the transmission quality of the useful information, but also enhances the strength of interference signals.

\begin{figure}[t]
	\centering
	\includegraphics[width=0.45\textwidth]{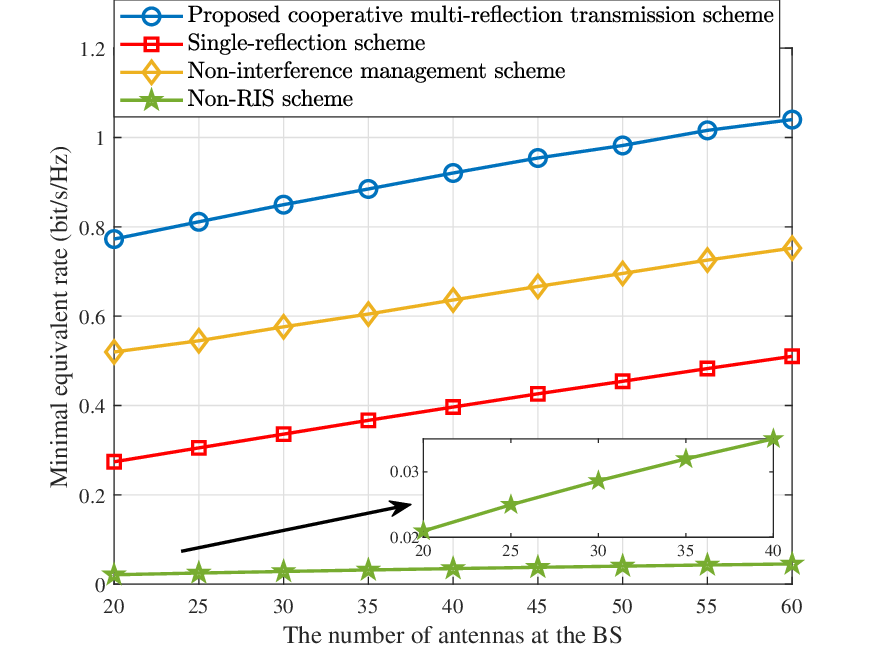}\\
\caption{The minimum equivalent rates for different numbers of antennas at the BS. The transmit power at the BS is $P_T=20$ dBW. There are $16$ RISs in the system, and the number of reflecting elements at each RIS is $M_j=20$, $j=1,2,...,16$.}
\label{fig:BS_antenna_increasing}
\end{figure}

\begin{figure}[t]
	\centering
\includegraphics[width=0.45\textwidth]{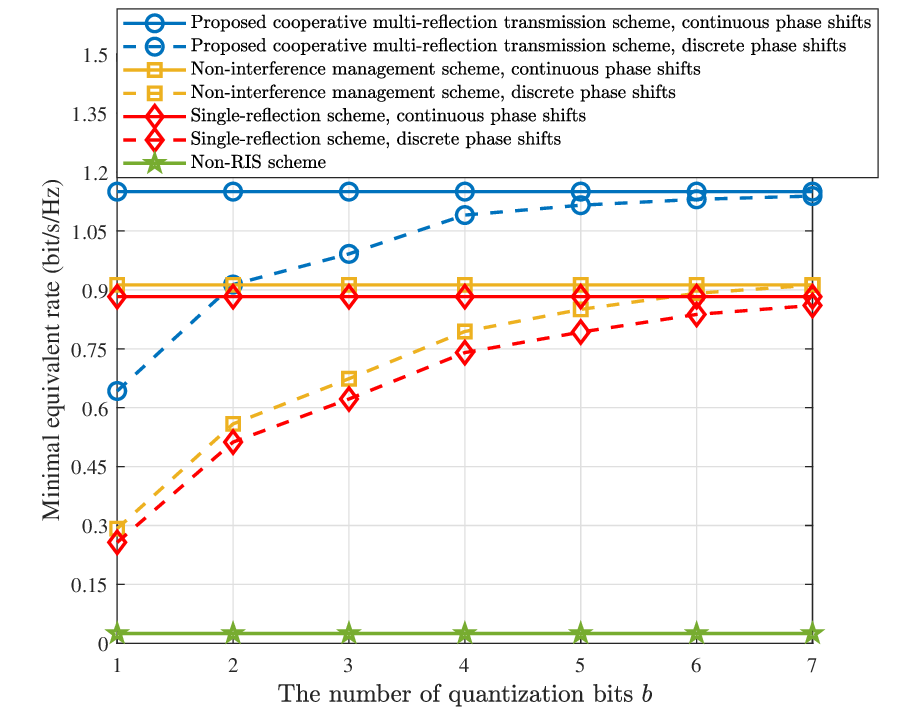}\\
\caption{The minimum equivalent data rates for different quantization bits $b$. The transmit power at the BS is $P_T=20$ dBW. There are $16$ RISs in the system, and the number of reflecting elements at each RIS is $M_j=20$, $j=1,2,...,16$.}
\label{fig:discrete}
\end{figure}

Fig. \ref{fig:BS_antenna_increasing} shows the minimum equivalent data rate of the proposed transmission scheme, single-reflection scheme, and non-RIS scheme, respectively, when the number of BS's antenna varies. In the simulation, we fix the transmit power at the BS as $P_T=20$ dBW, and each RIS has $M_j=20$ reflecting elements, $j=1,2,...,16$. We see from Fig. \ref{fig:BS_antenna_increasing} that due to the larger equivalent channel gain and well mitigated inter-user interference, our proposed multi-reflection transmission scheme achieves higher minimum data rate than the other benchmarks. Besides, we also find that the non-RIS scheme has the worst performance. This is because that no RIS is utilized during the signal transmission, thereby the channel gains at each user is much smaller than RIS-assisted schemes. Besides, since all the users communicate at the same time, the inter-user interference is quite large, resulting in poor performance.

Considering the deployment cost, it is possible to utilize discrete values for RISs' phase shifts in the practical implementations. In simulating the system, we refer to $b$ as the number of quantization bits and $B=2^b$ as the discrete phase shift level. Accordingly, the set of discrete phase-shift values at each element is
$
\mathcal{U}=\lbrace
0,\Delta\theta,...,(B-1)\Delta\theta
\rbrace,
$
where  $\Delta\theta=2\pi/B$. To show the impact of the discrete phase shift level on the minimum data rate performance, we depict the minimum data rate of all users versus the number of quantization bits $b$ in Fig. \ref{fig:discrete}. We observe from the results that as the number of quantization bits increases, the minimum data rate approaches that of the continuous case, and the proposed scheme can always achieve higher max-min rate under the same conditions. 

\vspace{-0.3cm}
	\section{Conclusions}
	In this paper, we investigate the cooperative multi-hop multi-RIS transmission framework for multi-user wireless communication systems. 
	With the aim to guarantee fairness among all users, we formulate an optimization problem to maximize the downlink minimum equivalent data rate of all users. This max-min equivalent rate problem comprehensively considers multi-reflection path selection, user grouping design, joint BS's beamforming optimization and group time allocation, which is difficult to solve. To effectively tackle it, we propose a novel three-step framework. We first analyzed the characteristics of the multi-RIS multi-reflection channels and reformulate the channels based on the composition features. With the reformulated channels, we derive the optimal RISs' phase shift design and assign corresponding weight to each link, with which we recast the multi-reflection path design as the shortest-path finding problem in graph theory, and select the optimal path with maximum channel gain for each user. We propose a new path separation constraint and divide users into different activation groups, so that each group takes turns in accessing the BS for signal reception during the scheduled time. In this way, severe inter-path inference caused by simultaneous transmissions is avoided. For more efficient time resource utilizing and generating higher system performance, we also propose a novel extension operation to allow users with less interference being included in different activation groups during the user grouping period. With obtained optimal multi-hop transmission path and user grouping schemes, we jointly optimize the BS' beamforming and user group scheduling to maximize the minimal equivalent data rate of all users by transforming the max-min problem into a feasibility check problem with the help of a series of auxiliary variables. The reformulated problem is finally solved by classical convex optimization approaches.
	Extensive simulations are carried out and the obtained simulation results have verified the superiority of the proposed multi-reflection transmission framework. 
	Useful insights into the optimal multi-reflection transmission design are drawn under different setups for multi-hop multi-RIS wireless communication systems.
	
	%%%%%%%%%%%%%%%%%%%%%%%%%%%%%%%%%%%%%%%%%%%%%%%%%%%%%%%%%%%%%%%%%%%%%%%%%%%%%%%%%%%	
	%	\appendices
	%	\section{Proof of Proposition 2}
	%	\label{Appendix_A}
	%	\hfill $\blacksquare$
	%%%%%%%%%%%%%%%%%%%%%%%%%%%%%%%%%%%%%%%%%%%%%%%%%%%%%%%%%%%%%%%%%%%%%%%%%%%%%%%%%%%%
	%\section{Proof of Proposition 3}
	%\label{Appendix_B}
	%\hfill $\blacksquare$
	%
	%%%%%%%%%%%%%%%%%%%%%%%%%%%%%%%%%%%%%%%%%%%%%%%%%%%%%%%%%%%%%%%%%%%%%%%%%%%%%%%%%%
	%\section{Convergence analysis of Algorithm 1}
	%\label{Appendix_D}
	%\hfill $\blacksquare$
	%%%%%%%%%%%%%%%%%%%%%%%%%%%%%%%%%%%%%%%%%%%%%%%%%%%%%%%%
	\begin{spacing}{1.232}
	\bibliographystyle{IEEEtran}
	\bibliography{myRef}
	\end{spacing}

\begin{IEEEbiography}[{\includegraphics[width=1in,height=1.25in,clip,keepaspectratio]{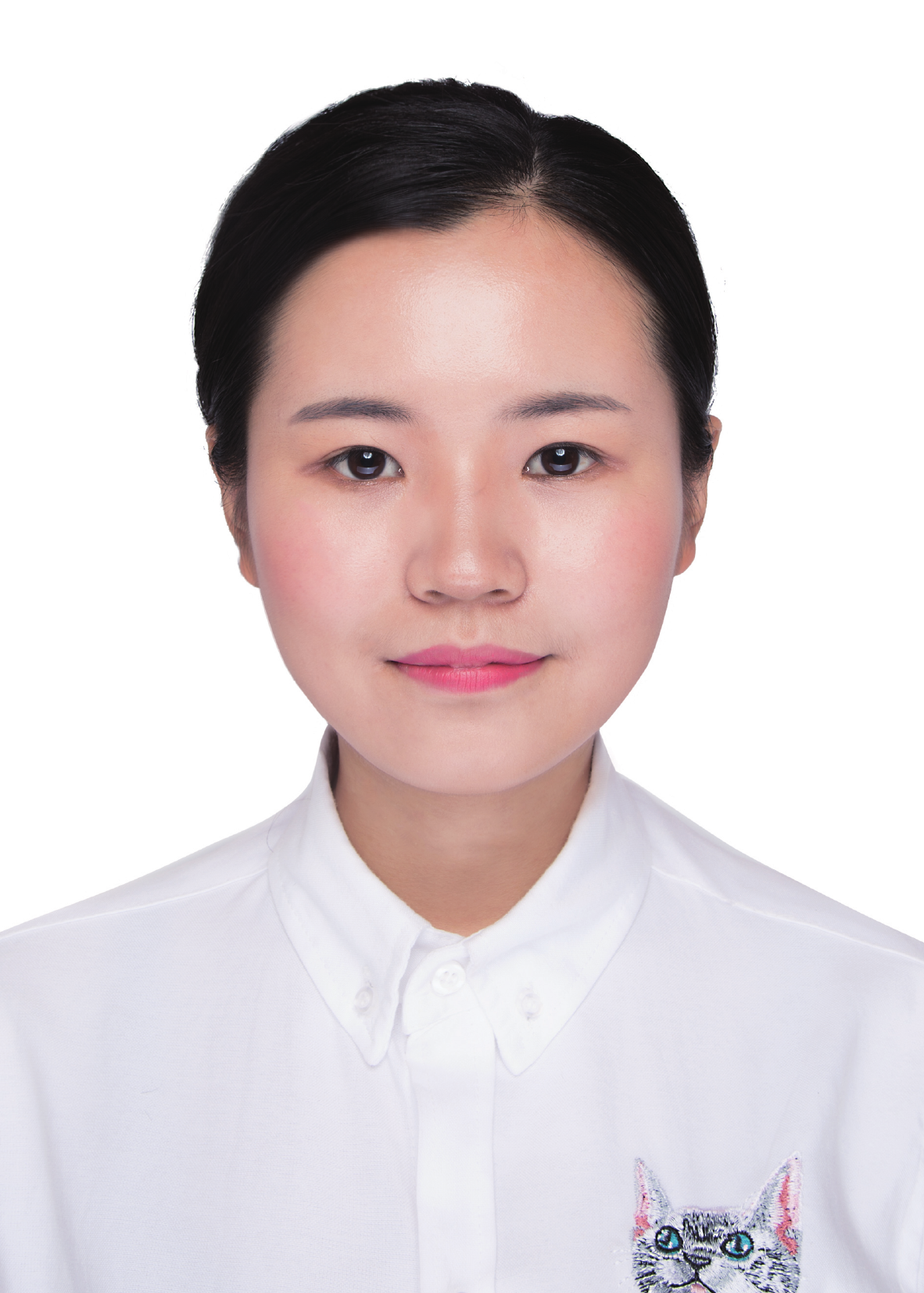}}]{Xiaoyan Ma} (Member, IEEE) received the B.S. and Ph.D. degrees from the School of Information Science and Engineering, Shandong University, China, in 2017 and 2023, respectively. She is currently a research fellow in School of Electrical and Electronic Engineering, Nanyang Technological University, Singapore. Her research interests include multiple-in multiple-out (MIMO) technology, reconfigurable intelligent surface (RIS) and signal processing technologies.
\end{IEEEbiography}

\begin{IEEEbiography}[{\includegraphics[width=1in,height=1.25in,clip,keepaspectratio]{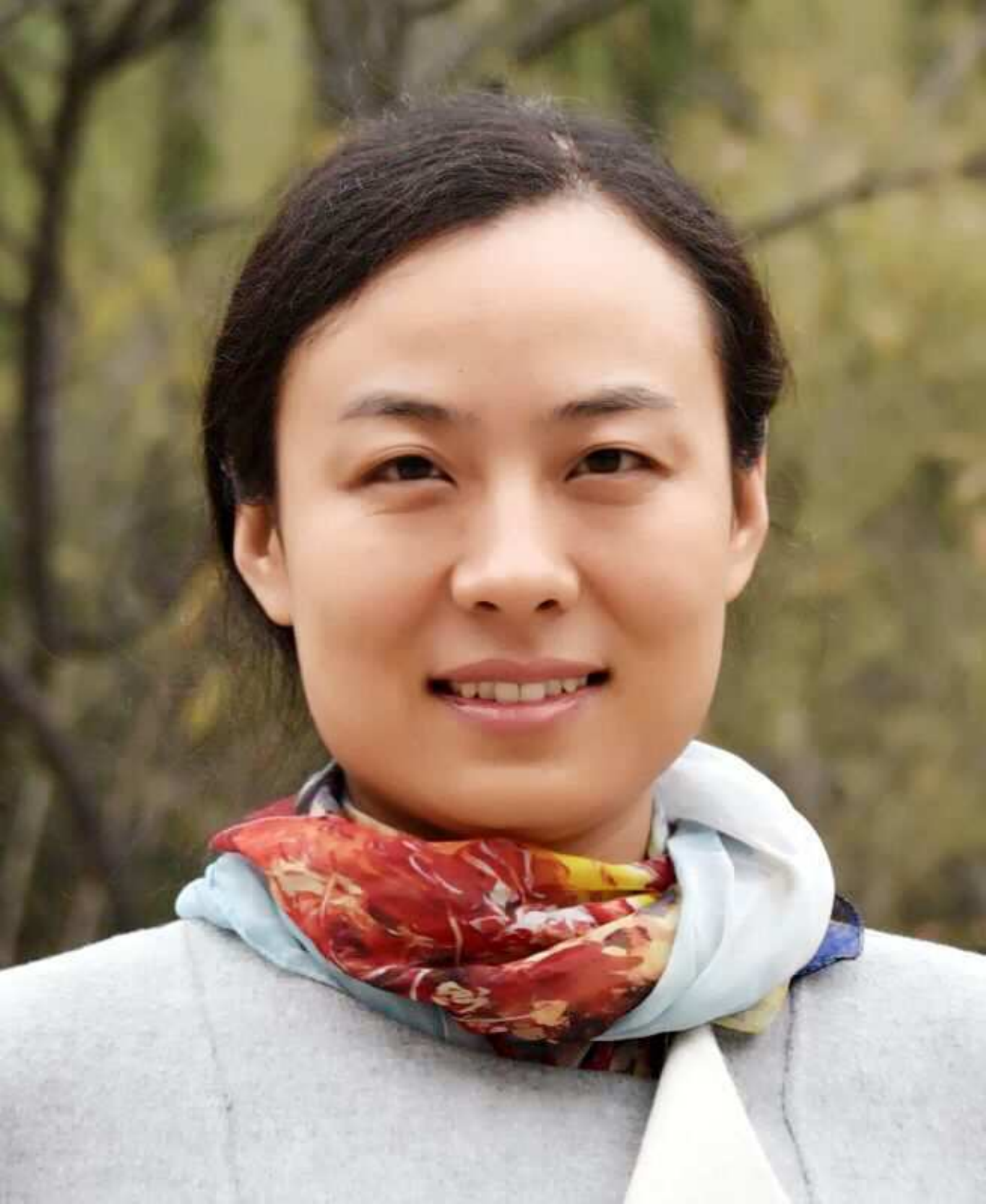}}]{Haixia Zhang} (Senior Member, IEEE) received the B.E. degree from the Department of Communication and Information Engineering, Guilin University of Electronic Technology, Guilin, China, in 2001, and the M.Eng. and Ph.D. degrees in communication and information systems from the School of Information Science and Engineering, Shandong University, Jinan, China, in 2004 and 2008, respectively. From 2006 to 2008, she was with the Institute for Circuit and Signal Processing, Munich University of Technology, Munich, Germany, as an Academic Assistant. From 2016 to 2017, she was a Visiting Professor with the University of Florida, Gainesville, FL, USA. She is currently a distinguished Professor with Shandong University. Dr. Zhang is actively participating in many professional services. She is/was an editor of the IEEE Transactions on Wireless Communications, IEEE Wireless Communication Letters, and China Communications and serves/served as Symposium Chairs, TPC Members, Session Chairs, and Keynote Speakers of many conferences. Her research interests include wireless communication and networks, industrial Internet of Things, wireless resource management, and mobile edge computing.
\end{IEEEbiography}

\begin{IEEEbiography}[{\includegraphics[width=1in,height=1.25in,clip,keepaspectratio]{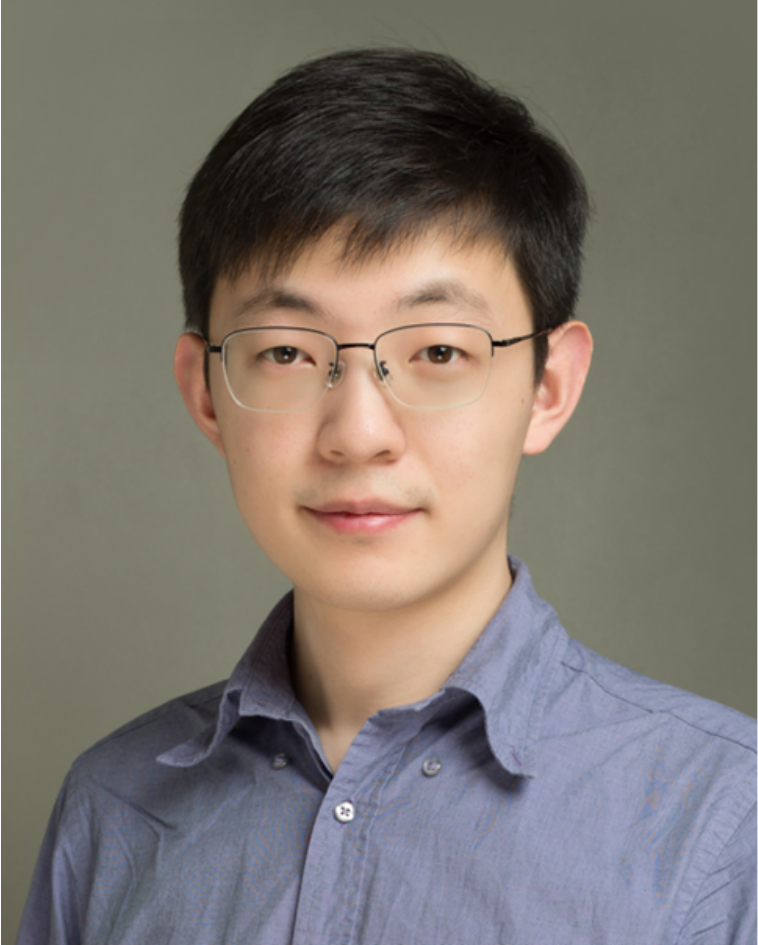}}]{Xianhao Chen} (Member, IEEE) received the B.Eng. degree in electronic information from Southwest Jiaotong University in 2017, and the Ph.D. degree in electrical and computer engineering from the University of Florida in 2022. He is currently an assistant professor with the Department of Electrical and Electronic Engineering, the University of Hong Kong. He received the 2022 ECE graduate excellence award for research from the University of Florida. His research interests include wireless networking, edge intelligence, and machine learning.
\end{IEEEbiography}

\begin{IEEEbiography}[{\includegraphics[width=1in,height=1.25in,clip,keepaspectratio]{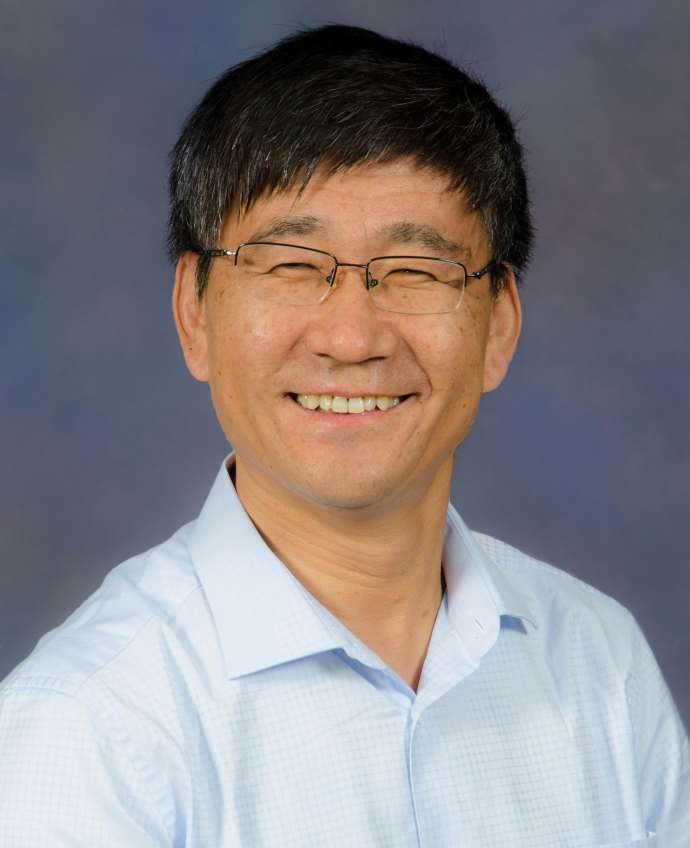}}]{Yuguang Fang} received an MS degree from Qufu Normal University, a PhD degree from Case Western Reserve University, and a PhD degree from Boston University. He joined the Department of Electrical and Computer Engineering at University of Florida in 2000 as an assistant professor, then was promoted to associate professor, full professor, and distinguished professor, respectively. Since 2022, he has been the Chair Professor of Internet of Things with Department of Computer Science at City University of Hong Kong. 

Dr. Fang received many awards including the US NSF CAREER Award, US ONR Young Investigator Award, 2018 IEEE Vehicular Technology Outstanding Service Award, IEEE Communications Society AHSN Technical Achievement Award, CISTC Technical Recognition Award, and WTC Recognition Award. He was the Editor-in-Chief of IEEE Transactions on Vehicular Technology and IEEE Wireless Communications and has served on several editorial boards of premier journals. He is a fellow of ACM, IEEE, and AAAS.  
\end{IEEEbiography}

\begin{IEEEbiography}[{\includegraphics[width=1in,height=1.25in,clip,keepaspectratio]{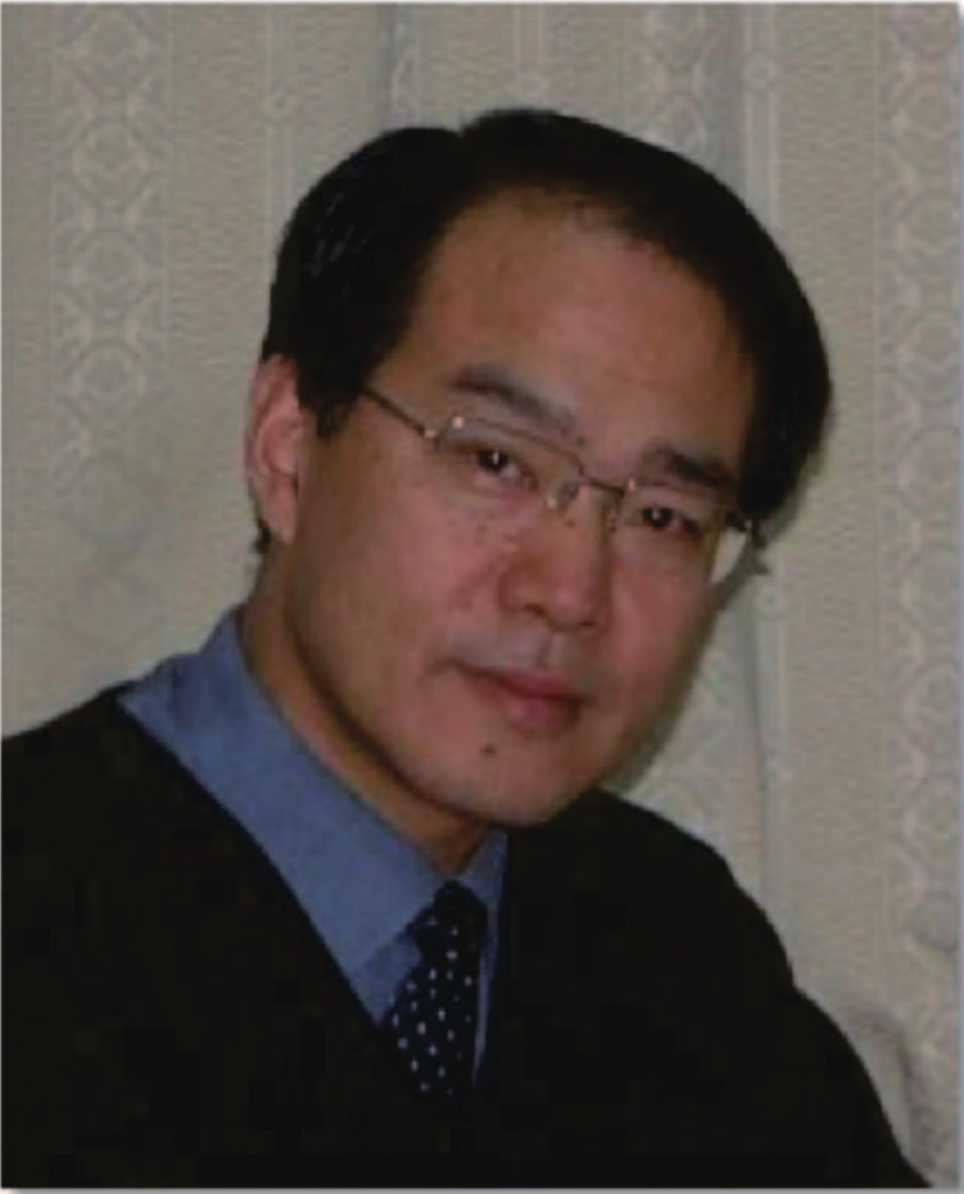}}]{Dongfeng Yuan}(SM'01) 
	received the M.S. degree from the Department of Electrical Engineering,
	Shandong University, China, in 1988, and received the Ph.D. degree from the Department of Electrical Engineering, Tsinghua University, China in January 2000. 
	Currently, he is a full professor in the School of Qilu Transportation, Shandong University, China. From 1993 to 1994, he was with the Electrical and Computer Department at the University of Calgary, Alberta, Canada. He was with the Department of Electrical Engineering in the University of Erlangen, Germany, from 1998 to 1999; with the Department of Electrical Engineering and Computer Science in the University of Michigan, Ann Arbor, USA, from 2001 to 2002; with the Department of Electrical Engineering in Munich University of Technology, Germany, in 2005; and with the Department of Electrical Engineering Heriot-Watt University, UK, in 2006.
	His current research interests include: Intelliegent communication systems, Mobile edge computing and cloud computing, AI and big data processing for communications.
\end{IEEEbiography}

\end{document}